\documentclass[runningheads,a4paper]{llncs}

\usepackage{amsmath,amssymb,dsfont}
\usepackage{graphicx,cancel,enumerate}
\usepackage{algorithm}
\usepackage[noend]{algpseudocode}
\usepackage{mwe}
\usepackage{subfig}
\usepackage[table]{xcolor}
\usepackage{float}
\usepackage{url}

\newtheorem{observation}[theorem]{Observation}

\newcommand{\rhob}{\ensuremath{\boldsymbol{\rho}}}
\newcommand{\deltab}{\ensuremath{\boldsymbol{\delta}}}

\algnewcommand\algorithmicinput{\textbf{Input:}}
\algnewcommand\INPUT{\item[\algorithmicinput]}

\algnewcommand\algorithmicoutput{\textbf{Output:}}
\algnewcommand\OUTPUT{\item[\algorithmicoutput]}

\begin{document}

\mainmatter  

\title{Density decompositions of networks}

\author{Glencora Borradaile\inst{1} \and
Theresa Migler\inst{2}\and
Gordon Wilfong\inst{3}}

\institute{School of EECS Oregon State University \and Cal Poly, San Luis Obispo \and Bell Labs}

\maketitle

\begin{abstract}
  We introduce a new topological descriptor of a network called the
  density decomposition which is a partition of the nodes of a network
  into regions of uniform density.  The decomposition we define is
  unique in the sense that a given network has exactly one density
  decomposition.  The number of nodes in each partition defines a
  density distribution which we find is measurably similar to the
  degree distribution of given {\em real} networks (social, internet,
  etc.) and measurably dissimilar in synthetic networks (preferential
  attachment, small world, etc.).
\end{abstract}

\section{Introduction}

A better understanding of the topological properties of real networks
can be advantageous for two major reasons.
First, knowing that a network has certain properties, e.g., bounded
degree or planarity, can sometimes allow for the design of more efficient 
algorithms for extracting information about the network or for the design of more
efficient distributed protocols to run on the network.
Second, it can lead to methods for synthesizing artificial networks that
more correctly match the properties of real networks thus allowing
for more accurate predictions of future growth of the network and more
accurate simulations of distributed protocols running on such a network.

We show that networks decompose naturally into regions of uniform
density, a {\em density decomposition}.  The decomposition we define
is unique in the sense that a given network has exactly one density
decomposition.  The number of nodes in each region defines a
distribution of the nodes according to the density of the region to
which they belong, that is, a {\em density distribution}
(Section~\ref{sec:density_sig}).  Although density is closely related
to degree, we find that the density distribution of a particular
network is not necessarily similar to the degree distribution of that
network.  For example, in many synthetic networks, such as those
generated by popular network models (e.g.\ preferential attachment and
small worlds), the density distribution is very different from the
degree distribution (Section~\ref{sec:dis}).  On the other hand, for
{\em all} of the real networks (social, internet, etc.) in our data
set, the density and degree distributions are measurably similar
(Section~\ref{sec:sim}).  Similar conclusions can be drawn using the
notion of {\em $k$-cores}~\cite{seidman1983}, but this suffers from
some drawbacks which we discuss in Section~\ref{sec:k-core}.

\subsection{Related work}
We obtain the density decomposition of a given undirected network
by first orienting the edges of this network in an {\em
  egalitarian}\footnote{An egalitarian orientation is one in which the
  indegrees of the nodes are as balanced as possible as allowed by the
  topology of the network.} manner. Then we partition the nodes
based on their indegree and connectivity in this orientation.

{\em Fair} orientations have been studied frequently in the
past. These orientations are motivated by many problems. One such
motivating problem is the following telecommunications network
problem: Source-sink pairs $(s_i,t_i)$ are linked by a directed
$s_i$-to-$t_i$ path $c_i$ (called a {\em circuit}).  When an edge of
the network fails, all circuits using that edge fail and must be
rerouted.  For each failed circuit, the responsibility for finding an
alternate path is assigned to either the source or sink corresponding
to that circuit. To limit the rerouting load of any vertex, it is
desirable to minimize the maximum number of circuits for which any
vertex is responsible.  Venkateswaran models this problem with an
undirected graph whose vertices are the sources and sinks and whose
edges are the circuits. He assigns the responsibility of a circuit's
potential failure by orienting the edge to either the source or the
sink of this circuit.  Minimizing the maximum number of circuits for
which any vertex is responsible can thus be achieved by finding an
orientation that minimizes the maximum indegree of any vertex.
Venkateswaran shows how to find such an
orientation~\cite{venkateswaran2004}. Asahiro, Miyano, Ono, and Zenmyo
consider the edge-weighted version of this problem~\cite{amoz2006}.
They give a combinatorial $\{{w_{max}\over
  w_{min}},(2-\epsilon)\}$-approximation algorithm where $w_{max}$ and
$w_{min}$ are the maximum and minimum weights of edges respectively,
and $\epsilon$ is a constant which depends on the input
\cite{amoz2006}. Klostermeyer considers the problem of reorienting
edges (rather than whole paths) so as to create graphs with given
properties, such as strongly connected graphs and acyclic graphs
\cite{klostermeyer99}. De Fraysseix and de Mendez show that they can
find an indegree assignment of the vertices given a particular
properties \cite{fm1994}. Biedl, Chan, Ganjali, Hajiaghayi, and Wood
give a $13\over 8$-approximation algorithm for finding an ordering of
the vertices such that for each vertex $v$, the neighbors of $v$ are
as evenly distributed to the right and left of $v$ as possible
\cite{bcghtw2005}.  For the purpose of deadlock
prevention~\cite{wimmer78}, Wittorff describes a heuristic for finding
an acyclic orientation that minimizes the sum over all vertices of the
function $\delta (v)$ choose $2$, where $\delta (v)$ is the indegree
of vertex $v$. This objective function is motivated by a problem
concerned with resolving deadlocks in communications
networks~\cite{Wittorff:2009:biblatex}.

In our work we show that the density decomposition can isolate the
densest subgraph. The densest subgraph problem has been studied a
great deal. Goldberg gives an algorithm to find the densest subgraph
in polynomial time using network flow techniques~\cite{goldberg1984}.
There is a 2-approximation for this problem that runs in linear time
\cite{charikar2000}. As a consequence of our decomposition, we find a
subgraph that has density no less than the density of the densest
subgraph less one. There are algorithms to find dense subgraphs in the
streaming model~\cite{bkv2012,gkt2005}. There are algorithms that find
all densest subgraphs in a graph (there could be many such
subgraphs)~\cite{shkrz2010}.

We consider many varied real networks in our study of the density
decomposition. We find our results to be consistent across biological,
technical, and social networks.

\section{The density decomposition}
\label{sec:density_sig}

In order to obtain the density decomposition of a given undirected
network we first orient the edges of this network in an egalitarian
manner. Then we partition the nodes based on their indegree and
connectivity in this orientation.

The following procedure, the {\sc Path-Reversal} algorithm, finds an
egalitarian orientation~\cite{bimowz2012}. A {\em reversible path} is
a directed path from a node $v$ to a node $u$ such that the indegree
of $v$, $\delta (v)$, is at least greater than the indegree of $u$
plus one: $\delta (v) > \delta (u) +1$
\begin{tabbing}
  \qquad \= Arbitrarily orient the edges of the network.\\
  \>  While there is a reversible path\\
  \> \qquad reverse this path.
\end{tabbing}
Since we are only reversing paths between nodes with differences in
indegree of at least 2, this procedure
converges; the running time of this algorithm is
quadratic~\cite{bimowz2012}.  The orientation resulting from this
termination condition suggests a hierarchical decomposition of its
nodes: Let $k$ be the maximum indegree in the orientation. {\em Ring}
$k$, denoted $R_k$, contains all nodes of indegree $k$ and all nodes
that reach nodes of indegree $k$. By the termination condition of the
above procedure, only nodes of indegree $k$ or $k -1$ are in $R_k$.
Iteratively, given $R_k, R_{k-1}, \ldots,$ and $R_{i+1}$, $R_i$
contains all the remaining nodes with indegree $i$ along with all the
remaining nodes that reach nodes with indegree $i$.  Nodes in $R_i$
must have indegree $i$ or $i-1$ by the termination condition of the
procedure.  By this definition, an edge between a node in $R_i$ and a
node in $R_j$ is directed from $R_i$ to $R_j$ when $i > j$ and all the
isolated nodes are in $R_0$.

Density can be defined in two ways: either as the ratio of number of edges
to number of nodes ($\frac{|E|}{|V|}$) or as the ratio of number of edges to
  total number of possible edges ($\frac{2|E|}{|V|(|V|-1)}$). In this
  discussion we use the former definition. This definition of density is
closely related to node degree (the number of edges adjacent to a
given node): the density of a network is equal to half the average
total degree.  

We {\em identify} a set $S$ of nodes in a graph by merging all
the nodes in $S$ into a single node $s$ and removing any self-loops
(corresponding to edges of the graph both of whose endpoints were in
$S$).
Our partition $R_k, R_{k-1}, \ldots, R_0$ induces regions of uniform
density in the following sense:

\begin{enumerate}[{\bf Density Property}]
\item  For any $i = 0, \ldots, k$, identifying the nodes in $\cup_{j > i}
  R_j$ and deleting the nodes in $\cup_{j < i} R_j$ leaves a network
  $G$ whose density is in the range $(i-1,i]$ (for $|R_i|$
  sufficiently large). 
\label{prop:density_prop}
\end{enumerate}

In particular, $R_k$ isolates a {\em densest} region in the
network.  
Consider the network $G_i$ formed by identifying the nodes $\cup_{j >
  i} R_j$ and deleting the nodes in $\cup_{j < i} R_j$; this network
has one node (resulting from identifying the nodes $\cup_{j > i} R_j$)
of indegree 0 and $|R_i|$ nodes of indegree $i$ of $i-1$, at least one
of which must have indegree $i$.  Therefore, for any $i$, the density
of $G_i$ is at most $i$ and density at least
\[\frac{(|R_i|-1)(i-1)+i}{|R_i|+1} \xrightarrow{|R_i| \gg i } i-1.\]

In Section~\ref{app:proof}, we observe that this relationship between
density and this decomposition is much stronger.


\subsection{Density and the Density Decomposition} \label{app:proof}
In this section we discuss the following three properties:

\begin{enumerate}[{\bf Property {D}1}]
\item The density of a densest subnetwork is at most
  $k$. That is, there is no denser region $R_j$
  for $j > k$.\label{prop:atmostk}
\item The density decomposition of a network is unique and does not
  depend on the starting orientation.\label{prop:unique}
\item Every densest subnetwork 
  contains only nodes of $R_k$. \label{prop:subring}
\end{enumerate}

These properties allow us to
unequivocally describe the density structure of a network.  We
summarize the density decomposition by the {\em density distribution}:
$(|R_0|,|R_1|,\dots |R_{k -1}|, |R_{k}|)$, i.e.~the number of
nodes in each region of uniform density.  We will refer to a node in
$R_i$ as having {\em density rank} $i$.

The subnetwork of a network $G$ {\em induced} by a subset $S$ of the nodes
of $G$ is defined as the set of nodes $S$ and the subset of edges of
$G$ whose endpoints are both in $S$; we denote this by $G[S]$.
First we will note  that both the densest subnetwork and the subnetwork
induced by the nodes of highest rank have density between $k-1$ and
$k$.  Recall that $k$ is the maximum indegree of a node in an egalitarian orientation of $G$ and that 
$R_i$ is the set of nodes in the $i^{th}$ ring of the density decomposition.  We will refer to $R_k$ as the densest ring.

We
use the following two lemmas to prove Property~D\ref{prop:atmostk}.

\begin{lemma}
\label{lem:den_top_ring}
  The density of the subnetwork induced by the nodes in $R_k$ is in the range $(k-1,k]$.
\end{lemma}


We could prove Lemma~\ref{lem:den_top_ring} directly with a simple
counting argument on the indegrees of nodes in $R_k$ (see Appendix~\ref{appendix:proofs}) or by using a
network flow construction similar to Goldberg's and the max flow-min
cut theorem~\cite{goldberg1984}.

\begin{lemma}
\label{lem:dense_between_k_k_1}
  The density of a densest subnetwork is in the range $(k-1,k]$.
\end{lemma}


The upper bound given in Lemma~\ref{lem:dense_between_k_k_1} may be
proven directly by using a counting argument for the indegrees of
vertices in an egalitarian orientation of the densest subnetwork (see Appendix~\ref{appendix:proofs}) or by
using the relationship between the density of the maximum density
subgraph and the psuedoarboricity~\cite{k2006}.

This upper bound proves Property~D\ref{prop:atmostk} of the density
decomposition. Property~D\ref{prop:atmostk} has been proven in another
context. It follows from a theorem of Frank and
Gy\'{a}rf\'{a}s~\cite{fg1976} that if $\ell$ is the maximum outdegree
in an orientation that minimizes the maximum outdegree then the
density of the network, $d$, is such that $\lceil d \rceil \leq \ell$.

\begin{corollary}
  The subgraph induced by the nodes of $R_k$ is at least as dense as
  the density of the densest subgraph less one.
\end{corollary}

Note that the partition of the rings does not rely on the
initial orientation, or, more strongly, nodes are uniquely partitioned into
rings, giving Property~D\ref{prop:unique}.

\begin{theorem}
  The density decomposition is unique. \label{thm:unique_rings}
\end{theorem}

We can prove this by noting that the maximum indegree of two
egalitarian orientations for a given network is the
same~\cite{bimowz2012,amoz2006,venkateswaran2004}. For a
contradiction, we consider two different egalitarian orientations of
the same graph that yield two distinct density decompositions. We then
compare corresponding rings in each orientation and find that they are
in fact the same (see Appendix~\ref{appendix:proofs}).

The following theorem relies on the fact that the density decomposition is unique and proves Property~D\ref{prop:subring}.

\begin{theorem}
  \label{thm:densest_contained_in_top_ring}
  The densest subnetwork of a network $G$ is induced by a subset of the nodes in the densest ring of $G$.
\end{theorem}

We could prove Theorem~\ref{thm:densest_contained_in_top_ring}
directly by comparing the density of the subgraph induced by the
vertices in the densest subgraph intersected with the vertices in
$R_k$ and the density of the densest subgraph (see Appendix~\ref{appendix:proofs}). Or we could use integer
parameterized max flow techniques~\cite{amoz2006,ggt1989}.

Note that there are indeed cases where the densest subgraph is induced
by a strict subset of nodes in the top ring. For example, consider the
graph, $G$, consisting of $K_3$ and $K_4$ with a single edge
connecting the two cliques. $K_4$ is the densest subgraph in $G$,
however all of $G$ is contained in the top ring ($R_2$).
 
\begin{figure}[h]
    \centering
    \includegraphics[scale=.8]{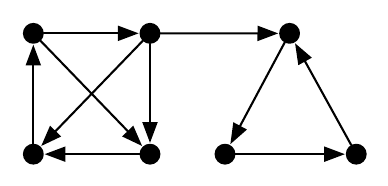}
    \caption{The orientation shown is an egalitarian orientation. In
      this graph all nodes are in the top ring. However, only the
      nodes in $K_4$ are in the densest subgraph.}
\label{fig:not_densest}
\end{figure}

See
Figure~\ref{fig:not_densest} for an example.

\subsection{Interpretation of density rank}\label{sec:rank}

\begin{figure}[h]
    \centering
    \includegraphics[scale=.8]{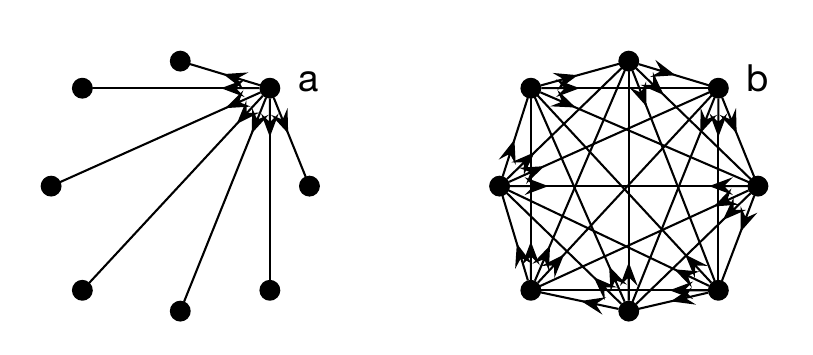}
    \label{fig:egalitarian_clique}
\caption{Two egalitarian orientations for networks with 8 nodes. This example generalizes to any number of nodes (Theorem~\ref{thm:indegree_clique}).}
\label{fig:egalitarian_orientations}
\end{figure}

We can interpret orientations as assigning responsibility: if an edge
is oriented from node $a$ to node $b$, we can view node $b$ as being
{\em responsible} for that connection.  Indeed several allocation
problems are modelled this
way~\cite{bimowz2012,ajmo2012,venkateswaran2004,amoz2006,hllt2003}.
Put another way, we can view a node as wishing to shirk as many of its
duties (modelled by incident edges) by assigning these duties to its
neighbors (by orienting the linking edge away from itself).  Of
course, every node wishes to shirk as many of its duties as possible.
However, the topology of the network may prevent a node from shirking
too many of its duties.  In fact, the egalitarian orientation is the
assignment in which every node is allowed to simultaneously shirk as
many duties as allowed by the topology of the network.  An example is
given in Figure~\ref{fig:egalitarian_orientations}; although nodes $a$
and $b$ both have degree 7, in the star network (left) $a$ can shirk all of
its duties, but in the clique network (right) $b$ can only shirk half of its
duties.  
There is a clear difference between these two cases that is
captured by the density rank of $a$ and $b$ that is not captured by the degree of $a$ and $b$.  For example, if these were co-authorship networks, the star network may represent a network in which author $a$ only co-authors papers with authors who never work with anyone else whereas the clique network shows that author $b$ co-authors with authors who also collaborate with others.  One may surmise that the work of author $b$ is more reliable or respected than the work of author $a$.

\begin{theorem}
  \label{thm:indegree_clique}
For a clique on $n$ nodes, there is an orientation where each
node has indegree either $\lfloor n/2 \rfloor$ or $\lfloor n/2
\rfloor -1$.
\end{theorem}


A proof for Theorem~\ref{thm:indegree_clique} can be given by
construction of such an egalitarian orientation (see
Appendix~\ref{appendix:proofs}) or by using a non-linear programming
approach~\cite{pq2006}.

\subsection{Relationship to $k$-cores} \label{sec:k-core}
 
A $k$-core of a network is the maximal subnetwork whose nodes all have
degree at least $k$~\cite{seidman1983}.  A $k$-core is found by
repeatedly deleting nodes of degree less than $k$ while possible.  For
increasing values of $k$, the $k$-cores form a nesting hierarchy (akin
to our density decomposition) of subnetworks $H_0,H_1,\ldots,H_p$
where $H_i$ is an $i$-core and $p$ is the smallest integer such that
$G$ has an empty $(p+1)$-core.  For networks generated by the
$G_{n,p}$ model, most nodes are in the
$p$-core~\cite{luczak1991,psw1996} For the preferential attachment
model, all nodes except the initial nodes belong to the $c$-core,
where $c$ is the number of edges connecting to each new
node~\cite{adbv2008}.  

These observations are similar to those we find for the density
distribution (Section~\ref{sec:sim}) and many of the observations we
make regarding the similarity of the degree and density distributions
of real networks also hold for $k$-core
decompositions~\cite{migler-dissertation}.  However, the local
definition of cores (depending only on the degree of a node) provides
a much looser connection to density than the density decomposition, as
we make formal in Lemma~\ref{lem:k-core-density}.  

The density of the top core may be less then the density of the top ring. Also, there are
graphs for which the densest subgraph is not contained in the top
core (see Appendix~\ref{append:k-cores}).

\begin{lemma}\label{lem:k-core-density}
  Given a core decomposition $H_0, H_1, \ldots, H_k$ of a network, the
  subnetwork formed by identifying the nodes in $\cup_{j > i}H_j$ and
  deleting the nodes in $\cup_{j< i} H_j$ has density in the range
  $[\frac{i}{2},i)$ for $|H_i|$ sufficiently large.
\end{lemma}

The proof of the above lemma can be found in Appendix~\ref{append:k-cores}.



\section{The similarity of degree and density distributions}
\label{sec:sim}

\begin{figure}[h!]
  \centering
  \includegraphics[scale=0.5]{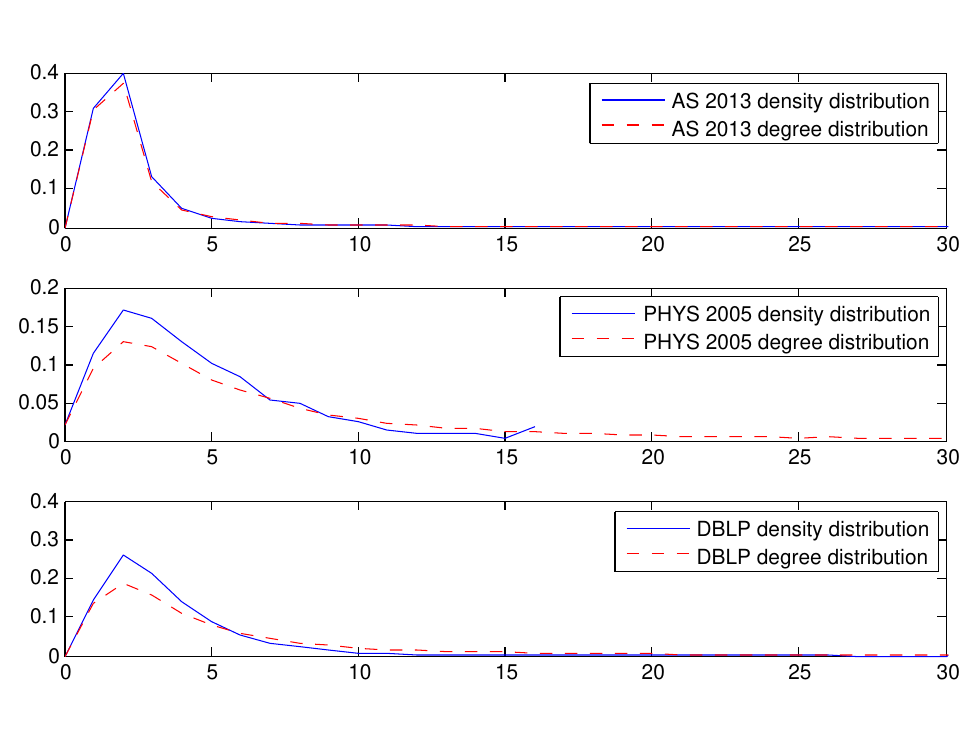}
  \caption{In the AS network nodes represent autonomous systems and
    two autonomous systems are connected if there is a routing
    agreement between them~\cite{as_website} (44,729 nodes and 170,735
    edges). In the PHYS network nodes represent condensed matter
    physicists and two physicists are connected if they have at least
    one co-authored paper~\cite{newman2004} (40,421 nodes and 175,692
    edges). In the DBLP network, nodes represent computer scientists
    and two computer scientists are connected if they have at least
    one co-authored paper~\cite{yl2012} (317,080 nodes and 1,049,866
    edges). The (truncated) normalized density and degree
    distributions are displayed.  The degree distributions have long
    diminishing tails. AS 2013 has 67 non-empty rings, but rings 31
    through 66 contain less than 1.5\% of the nodes; ring 67 contains
    0.75\% of the nodes.  DBLP has 4 non-empty rings denser than ring
    30 that are disconnected; rings 32, 40, 52 and 58 contain ~0.02\%,
    0.01\%, 0.03\% and 0.04\% of the nodes, respectively.}
  \label{fig:density_degree_real_networks}
\end{figure} 

\begin{figure}[h!]
  \centering
  \includegraphics[scale=0.5]{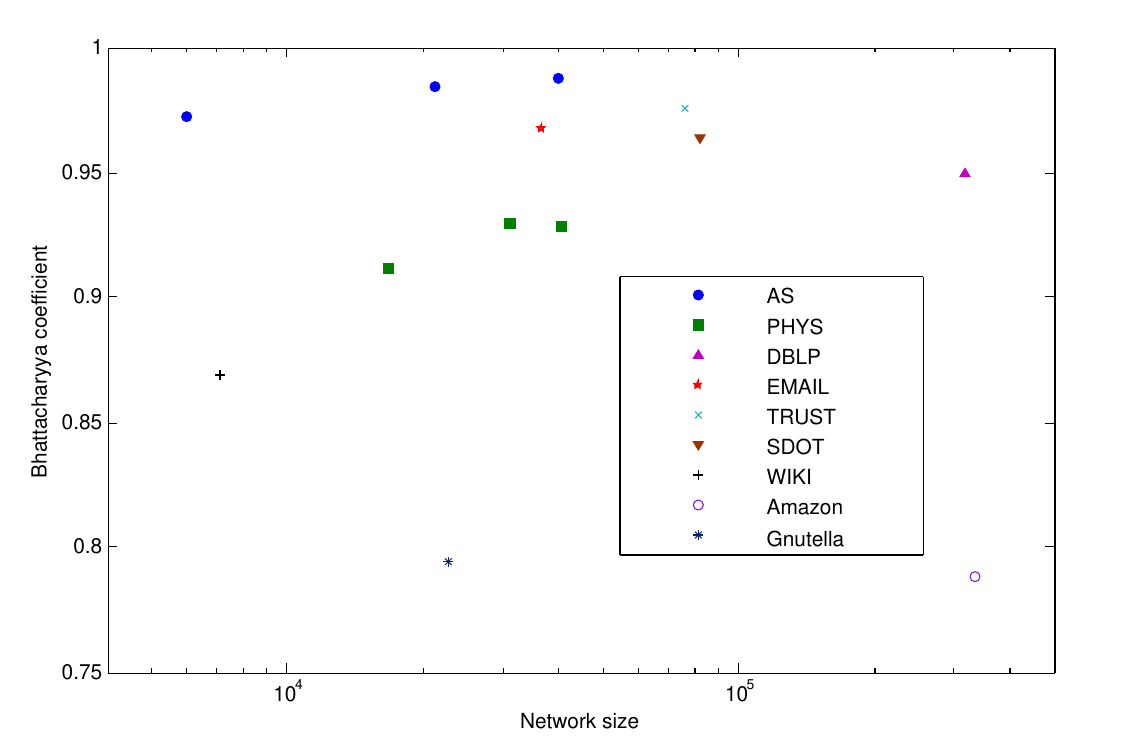}
  \caption{Similarity ($\beta_{\rho\delta}$) of density and degree
    distributions for 9 diverse networks. We introduced AS, PHYS, and
    DBLP in Figure~\ref{fig:density_degree_real_networks}. In the
    EMAIL network nodes represent Enron email addresses and two
    addresses are connected if there has been at least one email
    exchanged between them~\cite{ky2006} (36,692 nodes and 183,831
    edges). In the TRUST network nodes represent {\tt epinions.com}
    members and two members are connected if one trusts the
    other~\cite{tik2009} (75,879 nodes and 405,740 edges). In the SDOT
    network nodes represent {\tt slashdot.org} members and two members
    are connected if they are friends or foes~\cite{lldm2008} (82,168
    nodes and 504, 230 edges). In the WIKI network nodes represent
    {\tt wikipedia.org} users and two users are connected if one has
    voted for the other to be in an administrative role~\cite{lhk2012}
    (7,115 nodes and 103,689 edges). In the Amazon network nodes
    represent products and two products are connected if they are
    frequently purchased together~\cite{yl2012} (334,863 nodes and
    925,872 edges). In the Gnutella network nodes represent network
    hosts and two hosts are connected if they share
    files~\cite{rfi2002} (22,687 nodes and 54,704 edges). EMAIL ,
    TRUST, and WIKI are naturally directed networks. For these
    networks, we ignore direction and study the underlying undirected
    networks. Notice that both the Amazon and Gnutella networks are
    highly structured. It is not surprising that these networks would
    have a weaker connection between the density and degree
    distributions.}
  \label{fig:bc_real_networks}
\end{figure}

The normalized density $\rhob$ and degree $\deltab$ distributions for
three networks (AS 2013, PHYS 2005, and DBLP) are given in Figure~\ref{fig:density_degree_real_networks}, illustrating the similarity of the distributions.  We quantify the similarity between the density and degree distributions of these networks using the Bhattacharyya coefficient, $\beta$~\cite{bhattacharyya1943}.  For two normalized $\mathbf{p}$ and $\mathbf{q}$, the Bhattacharyya coefficient is: 
\[
\beta(\mathbf{p},\mathbf{q}) = \sum_i \sqrt{p_i\cdot q_i}.
\]
$\beta(\mathbf{p},\mathbf{q}) \in [0,1]$ for normalized, positive
distributions; $\beta(\mathbf{p},\mathbf{q}) = 0$ if and only if
$\mathbf{p}$ and $\mathbf{q}$ are disjoint;
$\beta(\mathbf{p},\mathbf{q}) = 1$ if and only if
$\mathbf{p}=\mathbf{q}$.  We denote the Bhattacharyya coefficient comparing the normalized density $\rhob$ and degree $\deltab$ distributions, $\beta(\rhob,\deltab)$ for a network $G$ by $\beta_{\rho\delta}(G)$.  Specifically,
\[
\beta_{\rho\delta}(G) = \beta(\rhob,\deltab) = \sum_i \sqrt{\rho_i\cdot \delta_i}
\]
where $\rho_i$ is the fraction of nodes in the $i^{th}$ ring of the
density decomposition of $G$ and $\delta_i$ is the fraction of nodes
of {\em total} degree $i$ in $G$; we take $\rho_i = 0$ for $i > k$
where $k$ is the maximum ring index. Refer to
Figure~\ref{fig:bc_real_networks}.  For all the networks in our data
set, $\beta_{\rho\delta} > 0.78$.  Note that if we exclude the
Gnutella and Amazon networks, $\beta_{\rho\delta}>0.9$.  We point out
that the other networks are self-determining in that each relationship
is determined by at least one of the parties involved.  On the other
hand, the Gnutella network is highly structured and designed and the
Amazon network is a is a one-mode projection of the buyer-product
network (which is in turn self-determining).

Perhaps this is not surprising, given the close relationship between
density and degree; one may posit that the density distribution
$\rhob$ simply bins the degree distribution $\deltab$.  However, note
that a node's degree is its {\em total} degree in the undirected
graph, whereas a node's rank is within one of its {\em indegree} in an
egalitarian orientation.  Since the total indegree to be shared
amongst all the nodes is half the total degree of the network, we
might assume that, if the density distribution is a binning of the
degree distribution, the density rank of a node of degree $d$ would be
roughly $d/2$.  That is, we may expect that the density distribution
is halved in range and doubled in magnitude ($\rho_i \approx
2{\delta}_{2i}$).  If this is the case, then
\[\beta({\rhob},\deltab) \approx \sum_d \sqrt{\rho_i\delta_i} \approx \sum_d \sqrt{2\delta_d{\delta}_{2d}}.\]
If we additionally assume that our network has a power-law degree
distribution such as $\delta_x \varpropto 1 /
x^3$,
\[\beta({\rhob},{\deltab}) \approx \int_1^\infty
\sqrt{\frac{2}{x^3}\left(2 {\frac{2}{(2x)^3}}\right)} \mathrm{d}x =
0.5\]
(after normalizing the distributions and using a continuous
approximation of $\beta$).  Even with these idealized assumptions,
this does not come close to explaining $\beta_{\rho\delta}$ being in
excess of 0.78 for the networks in our data set.  Further, for
many synthetic networks $\beta_{\rho\delta}$ is small, as we
discuss in the next section.  We note that this separation between
similarities of density and degree distributions for the empirical
networks and synthetic networks can be illustrated with almost any
divergence or similarity measure for a pair of distributions.

\subsection{The dissimilarity of degree and density distributions of random networks}
\label{sec:dis}
In contrast to the measurably similar degree and density distributions
of real networks, the degree and density distributions are measurably
{\em dis}similar for networks produced by many common random network
models; including the preferential attachment (PA) model of Barabasi
and Albert~\cite{BA99} and the small world (SW) model of Watts and
Strogatz~\cite{ws1998}.  We use $\tilde \beta_ {\rho\delta} (M)$ to
denote the Bhattacharyya coefficient comparing the expected degree and
density distributions of a network generated by a model $M$.

\paragraph{Preferential attachment networks} In the PA model, a small
number, $n_0$, of nodes seed the network and nodes are added
iteratively, each attaching to a fixed number, $c$, of existing nodes.
Consider the orientation where each added edge is directed toward the
newly added node; in the resulting orientation, all but the $n_0$ seed
nodes have indegree $c$ and the maximum indegree is $c$.  At most
$cn_0$ path reversals will make this orientation egalitarian, and,
since $cn_0$ is typically very small compared to $n$ (the total number
of nodes), most of the nodes will remain in the densest ring $R_c$.
Therefore PA networks have nearly-trivial density distributions:
$\rho_c \approx 1$. On the other hand the expected fraction of degree
$c$ nodes is $2/(c+2)$~\cite{barabasi2016}.  Therefore
$\tilde \beta_{\rho\delta}(\mbox{PA})\approx \sqrt{2/(c+2)}$.

\paragraph{Small-world networks} A small-world network is one
generated from a $d$-regular network
by reconnecting (uniformly at random) at least
one endpoint of every edge with some probability.  For probabilities
close to 0, a network generated in this way is close to $d$-regular;
for probabilities close to 1, a network generated this way approaches
one generated by the random-network model ($G_{n,p}$) of Erd\"os and
R\'enyi~\cite{er1959}. In the first extreme,
$\tilde\beta_{\rho\delta}(SW) = 0$ (Lemma~\ref{cl:d-reg} below)
because all the nodes have the same degree and the same rank.  As the
reconnection probability increases, nodes are not very likely to
change rank while the degree distribution spreads slightly.  In the
second extreme, the highest rank of a node is
$\lfloor c/2 \rfloor +1$~\cite{gps2014} and, using an observation of
the expected size of the densest subnetwork
\footnote{e-mail exchange between Glencora Borradaile and Abbas
  Mehrabian}, with high probability nearly all the nodes have this
rank. It follows that
\[
\tilde \beta_{\rho\delta}(G_{n,p}) \approx \sqrt{\frac{c^{c/2}}{e^{-c}(c/2)!}}, \]
which approaches 0 very quickly as $c$ grows.  We verified this
experimentally finding that $\tilde
\beta_{\rho\delta}(G_{n,p}) < 0.5$ for $c \ge 5$.

\begin{lemma}\label{cl:d-reg}
  For $d \ge 3$, $\beta_{\rho\delta}(G) = 0$ for any $d$-regular network $G$ with $d \ge 3$.
\end{lemma}

\begin{proof}
  We argue that $\rho_d = 0$, proving the lemma since $\delta_d = 1$ for a $d$-regular network.
  For a contradiction, suppose $\rho_d > 0$.  Then $|R_d| = x$ for some $x > 0$, where $R_d$ is the
  set of nodes of $G$ in the $d^{th}$ ring of $G$'s density
  decomposition.  Note that the highest rank node in $G$ has rank at
  most $d$, since there are no nodes with degree $> d$.  Let $H$ be
  the subnetwork of $H$ containing all the nodes of $R_d$ and all
  the edges of $G$ both of whose endpoints are in $R_d$.  $H$ has at
  least one node of indegree $d$ and all other nodes have indegree at
  least $d-1$; therefore $H$ must have at least $d+(x-1)(d-1)$ edges.  On
  the other hand, the total degree of every node in $H$ is at most
  $d$, so $H$ has at most $dx/2$ edges.  We must have
  $d+(x-1)(d-1)\le dx/2$, which is a contradiction for $d \ge 3$ and $x > 0$.
\end{proof}

\section{Conclusion}
We have introduced the density decomposition and summarized the
decomposition with the density distribution. We found that the
hierarchy of vertices within this decomposition are partitioned according to
the density of the induced subgraphs. We found that the density and degree
distributions are remarkably similar in real graphs and dissimilar in
synthetic networks. In the appendix we discuss using the density
distribution to build more realistic random graph models.

\bibliographystyle{plain}
\bibliography{density_decomposition}

\begin{thebibliography}{10}

\bibitem{adbv2008}
Jose Alvarez-Hamelin, Luca Dall’Asta, Alain Barrat, and Alessandro
  Vespignani.
\newblock k-core decomposition of {I}nternet graphs: hierarchies,
  self-similarity and measurement biases.
\newblock {\em Networks and Heterogeneous Media}, 3(2):371, 2008.

\bibitem{ajmo2012}
Yuichi Asahiro, Jesper Jansson, Eiji Miyano, and Hirotaka Ono.
\newblock Upper and lower degree bounded graph orientation with minimum
  penalty.
\newblock In {\em Proceedings in Computing: The Australasian Theory Symposium},
  pages 139--146, 2012.

\bibitem{amoz2006}
Yuichi Asahiro, Eiji Miyano, Hirotaka Ono, and Kouhei Zenmyo.
\newblock Graph orientation algorithms to minimize the maximum outdegree.
\newblock {\em International Journal of Foundations of Computer Science},
  18(2):197--215, 2007.

\bibitem{bkv2012}
Bahman Bahmani, Ravi Kumar, and Sergei Vassilvitskii.
\newblock Densest subgraph in streaming and mapreduce.
\newblock {\em Proc. VLDB Endow.}, 5(5):454--465, January 2012.

\bibitem{barabasi2016}
Albert-L\'{a}szl\'{o} Barab\'{a}si.
\newblock {\em Network Science}.
\newblock Cambridge University Press, Cambridge, 2016.

\bibitem{BA99}
Albert-L\'{a}szl\'{o} Barab\'{a}si and R\'{e}ka Albert.
\newblock Emergence of scaling in random networks.
\newblock {\em Science}, 286(5439):509--512, 1999.

\bibitem{bz2003}
Vladimir Batagelj and Matjaz Zaversnik.
\newblock An {O(m)} algorithm for cores decomposition of networks.
\newblock {\em Advances of Data Analysis and Classification}, 5:129--145, 2011.

\bibitem{bhattacharyya1943}
Anil~Kumar Bhattacharyya.
\newblock On a measure of divergence between two statistical populations
  defined by their probability distributions.
\newblock {\em Bull. Calcutta Math. Soc.}, 35:99--109, 1943.

\bibitem{bb2001}
Ginestra Bianconi and Albert-Laszlo Barabási.
\newblock Competition and multiscaling in evolving networks.
\newblock {\em EPL (Europhysics Letters)}, 54(4):436, 2001.

\bibitem{bcghtw2005}
Therese Biedl, Timothy Chan, Yashar Ganjali, Mohammad~Taghi Hajiaghayi, and
  David~R. Wood.
\newblock Balanced vertex-orderings of graphs.
\newblock {\em Discrete Appl. Math.}, 148:27--48, April 2005.

\bibitem{bimowz2012}
Glencora Borradaile, Jennifer Iglesias, Theresa Migler, Antonio Ochoa, Gordon
  Wilfong, and Lisa Zhang.
\newblock Egalitarian graph orientations.
\newblock In {\em Journal of Graph Algorithms and Applications}, volume~21,
  pages 687--708, 2017.

\bibitem{charikar2000}
Moses Charikar.
\newblock Greedy approximation algorithms for finding dense components in a
  graph.
\newblock In {\em Proceedings of the Third International Workshop on
  Approximation Algorithms for Combinatorial Optimization}, pages 84--95,
  London, UK, 2000. Springer-Verlag.

\bibitem{er1959}
Paul Erd\"{o}s and Alfr\'{e}d R\'{e}nyi.
\newblock On random graphs, {I}.
\newblock {\em Publicationes Mathematicae (Debrecen)}, 6:290--297, 1959.

\bibitem{fg1976}
Andr\'{a}s Frank and A.~Gy\'{a}rf\'{a}s.
\newblock How to orient the edges of a graph?
\newblock {\em Colloquia Mathematica Societatis J\'{a}nos Bolyai}, 1:353--364,
  1976.

\bibitem{fm1994}
Hubert~de Fraysseix and Patrice~Ossona de~Mendez.
\newblock Regular orientations, arboricity, and augmentation.
\newblock In {\em Proceedings of the DIMACS International Workshop on Graph
  Drawing}, pages 111--118, London, UK, 1995. Springer-Verlag.

\bibitem{ggt1989}
Giorgio Gallo, Michael Grigoriadis, and Robert Tarjan.
\newblock A fast parametric maximum flow algorithm and applications.
\newblock {\em SIAM Journal on Computing}, 18(1):30--55, February 1989.

\bibitem{gkt2005}
David Gibson, Ravi Kumar, and Andrew Tomkins.
\newblock Discovering large dense subgraphs in massive graphs.
\newblock In {\em Proceedings of the 31st international conference on very
  large data bases}, pages 721--732. VLDB Endowment, 2005.

\bibitem{goldberg1984}
Andrew Goldberg.
\newblock Finding a maximum density subgraph.
\newblock Technical report, University of California at Berkeley, Berkeley, CA,
  USA, 1984.

\bibitem{hllt2003}
Nicholas J.~A. Harvey, Richard~E. Ladner, L\'{a}szl\'{o} Lov\'{a}sz, and Tami
  Tamir.
\newblock Semi-matchings for bipartite graphs and load balancing.
\newblock {\em Journal of Algorithms}, 59:53--78, 2006.

\bibitem{js1998}
Mark Jerrum and Gregory~B. Sorkin.
\newblock The metropolis algorithm for graph bisection.
\newblock {\em Discrete Applied Mathematics}, 82(1–3):155 -- 175, 1998.

\bibitem{kv2003}
Jeong~Han Kim and Van~H. Vu.
\newblock Generating random regular graphs.
\newblock In {\em Proceedings of the thirty-fifth annual ACM symposium on
  Theory of computing}, STOC '03, pages 213--222, New York, NY, USA, 2003. ACM.

\bibitem{ky2006}
Bryan Klimt and Yiming Yang.
\newblock Introducing the {E}nron {C}orpus.
\newblock In {\em First Conference on Email and Anti-Spam}, 2004.

\bibitem{klostermeyer99}
William~F. Klostermeyer.
\newblock Pushing vertices and orienting edges.
\newblock {\em Ars Combinatorial}, 51:65--75, 1999.

\bibitem{k2006}
\L{}ukasz Kowalik.
\newblock Approximation scheme for lowest outdegree orientation and graph
  density measures.
\newblock In {\em Proceedings of the 17th International Conference on
  Algorithms and Computation}, ISAAC'06, pages 557--566, Berlin, Heidelberg,
  2006. Springer-Verlag.

\bibitem{ls2009}
Silvio Lattanzi and D.~Sivakumar.
\newblock Affiliation networks.
\newblock In {\em Proceedings of the Forty-first Annual ACM Symposium on Theory
  of Computing}, STOC '09, pages 427--434, New York, NY, USA, 2009. ACM.

\bibitem{lhk2012}
Jure Leskovec, Daniel Huttenlocher, and Jon Kleinberg.
\newblock Signed networks in social media.
\newblock In {\em Proceedings of the SIGCHI Conference on Human Factors in
  Computing Systems}, CHI '10, pages 1361--1370, New York, NY, USA, 2010. ACM.
\newblock http://snap.stanford.edu/data/.

\bibitem{lkf2005}
Jure Leskovec, Jon Kleinberg, and Christos Faloutsos.
\newblock Graphs over time: Densification laws, shrinking diameters and
  possible explanations.
\newblock In {\em Proceedings of the eleventh ACM SIGKDD International
  Conference on Knowledge Discovery in Data Mining}, pages 177--187. ACM Press,
  2005.

\bibitem{lldm2008}
Jure Leskovec, Kevin Lang, Anirban Dasgupta, and Michael Mahoney.
\newblock Community structure in large networks: Natural cluster sizes and the
  absence of large well-defined clusters.
\newblock {\em CoRR}, abs/0810.1355, 2008.
\newblock http://snap.stanford.edu/data/.

\bibitem{luczak1991}
Tomasz \L{}uczak.
\newblock Size and connectivity of the k-core of a random graph.
\newblock {\em Discrete Mathematics}, 91(1):61--68, August 1991.

\bibitem{mb1983}
David~W. Matula and Leland~L. Beck.
\newblock Smallest-last ordering and clustering and graph coloring algorithms.
\newblock {\em J. ACM}, 30(3):417--427, July 1983.

\bibitem{ml2014}
Julian Mcauley and Jure Leskovec.
\newblock Discovering social circles in ego networks.
\newblock {\em ACM Transactions on Knowledge Discovery from Data},
  8(1):4:1--4:28, February 2014.

\bibitem{mw1990}
Brendan~D. McKay and Nicholas~C. Wormald.
\newblock Uniform generation of random regular graphs of moderate degree.
\newblock {\em J. Algorithms}, 11(1):52--67, February 1990.

\bibitem{migler-dissertation}
Theresa Migler.
\newblock {\em The Density Signature}.
\newblock PhD thesis, Oregon State University, 2014.

\bibitem{newman2004}
Mark Newman.
\newblock Fast algorithm for detecting community structure in networks.
\newblock {\em Phys. Rev. E}, 69:066133, Jun 2004.
\newblock http://www-personal.umich.edu/\~{}mejn/netdata/.

\bibitem{newman2010}
Mark Newman.
\newblock {\em Networks: An Introduction}.
\newblock Oxford University Press, Inc., New York, NY, USA, 2010.

\bibitem{pq2006}
Jean-Claude Picard and Maurice Queyranne.
\newblock A network flow solution to some nonlinear 0-1 programming problems,
  with applications to graph theory.
\newblock {\em Networks}, 12:141--159, 1982.

\bibitem{psw1996}
Boris Pittel, Joel Spencer, and Nicholas Wormald.
\newblock Sudden emergence of a giant k-core in a random graph.
\newblock {\em J. Comb. Theory Ser. B}, 67(1):111--151, May 1996.

\bibitem{rfi2002}
Matei Ripeanu, Ian Foster, and Adriana Iamnitchi.
\newblock Mapping the {G}nutella network: {P}roperties of large-scale
  peer-to-peer systems and implications for system design.
\newblock {\em IEEE Internet Computing Journal}, 6:2002, 2002.
\newblock http://snap.stanford.edu/data/.

\bibitem{shkrz2010}
Barna Saha, Allison Hoch, Samir Khuller, Louiqa Raschid, and Xiao-Ning Zhang.
\newblock Dense subgraphs with restrictions and applications to gene annotation
  graphs.
\newblock In {\em Proceedings of the 14th Annual international conference on
  Research in Computational Molecular Biology}, RECOMB'10, pages 456--472,
  Berlin, Heidelberg, 2010. Springer-Verlag.

\bibitem{seidman1983}
Stephen Seidman.
\newblock Network structure and minimum degree.
\newblock {\em Social Networks}, 5(3):269--287, 1983.

\bibitem{tik2009}
Maryam Tahajod, Azadeh Iranmehr, and Nasim Khozooyi.
\newblock Trust management for semantic web.
\newblock In {\em Computer and Electrical Engineering, 2009. ICCEE '09. Second
  International Conference}, volume~2, pages 3--6, 2009.
\newblock http://snap.stanford.edu/data/.

\bibitem{venkateswaran2004}
Venkat Venkateswaran.
\newblock Minimizing maximum indegree.
\newblock {\em Discrete Appl. Math.}, 143:374--378, September 2004.

\bibitem{vl2005}
Fabien Viger and Matthieu Latapy.
\newblock Efficient and simple generation of random simple connected graphs
  with prescribed degree sequence.
\newblock In {\em The Eleventh International Computing and Combinatorics
  Conference}, pages 440--449. Springer, 2005.

\bibitem{ws1998}
Duncan Watts and Steven Strogatz.
\newblock Collective dynamics of `small-world' networks.
\newblock {\em Nature}, 393(6684):409--10, 1998.

\bibitem{wimmer78}
W.~Wimmer.
\newblock {E}in {V}erfahren zur {V}erhinderung von {V}erklemmungen in
  {V}ermittlernetzen.
\newblock
  \url{http://www.worldcat.org/title/verfahren-zur-verhinderung-von-verklemmungen-in-vermittlernetzen/},
  October 1978.

\bibitem{Wittorff:2009:biblatex}
Vaughan Wittorff.
\newblock Implementation of constraints to ensure deadlock avoidance in
  networks, 2009.
\newblock US Patent \# 7,532,584 B2.

\bibitem{gps2014}
Pu~Gao Xavier P\'{e}rez-Gim\'enez and Cristiane Sato.
\newblock Arboricity and spanning-tree packing in random graphs with an
  application to load balancing.
\newblock In {\em Proceedings of the Twenty-Fifth Annual ACM-SIAM Symposium on
  Discrete Algorithms}, SODA '14, pages 317--326. SIAM, 2014.

\bibitem{yl2012}
Jaewon Yang and Jure Leskovec.
\newblock Defining and evaluating network communities based on ground-truth.
\newblock In {\em Proceedings of the ACM SIGKDD Workshop on Mining Data
  Semantics}, MDS '12, pages 3:1--3:8, New York, NY, USA, 2012. ACM.
\newblock http://snap.stanford.edu/data/.

\bibitem{as_website}
Yu~Zhang.
\newblock {I}nternet {AS}-level {T}opology {A}rchive.
\newblock http://irl.cs.ucla.edu/topology/.

\end{thebibliography}

\newpage
\appendix

\section{Proofs}\label{appendix:proofs}
In this section we include proofs for theorems in
Sections~\ref{app:proof} and~\ref{sec:rank}.

\begin{lemma}
\label{lem:den_top_ring}
  The density of the subnetwork induced by the nodes in $R_k$ is in the range $(k-1,k]$.
\end{lemma}

\begin{proof}  
  All nodes in $R_k$ have indegree $k$ or $k-1$ in $G$. Since any edge
  incident to a node in $R_k$ but not in $G[R_k]$ is directed out of
  $R_k$ in $G$, the indegree of every node in $G[R_k]$ is $k$ or
  $k-1$.  Let $n_k$ be the number of nodes of indegree $k$ in $G[R_k]$
  and $n_{k-1}$ be the number of nodes of degree $k-1$ in
  $G[R_{k-1}]$.  Therefore, the number of edges in $G[R_k]$ is $kn_k +
  (k-1)n_{k-1}$ and: 
  \[
  \text{density}(G[R_k]) = {\frac{kn_k + (k-1)n_{k-1}}{n_k +n_{k-1}}} \le k
  \]
  Since there is at least one node of indegree $k$ in $G[R_k]$, $n_k > 0$.  Therefore:
  \[
  {\frac{kn_k + (k-1)n_{k-1}}{n_k +n_{k-1}}} = \frac{(k-1)(n_k + n_{k-1})+n_k}{n_k +n_{k-1}}> k-1
  \]
\end{proof}

\begin{lemma}
\label{lem:dense_between_k_k_1}
  The density of a densest subnetwork is in the range $(k-1,k]$.
\end{lemma}

\begin{proof}
  Let $H$ be a densest subnetwork and let each edge of $H$ inherit
  the orientation of the same edge in an egalitarian orientation of
  $G$.  Every node of $H$ has indegree at most $k$ (when restricted to $H$).  Therefore 
  \[
  \text{density}(H) \le \frac{n_H k}{n_H}\le k
  \]
  where $n_H$ is the number of nodes in $H$.
 Furthermore, by Lemma~\ref{lem:den_top_ring}, the density of
  $G[R_k]$ is greater than $k-1$ and so the densest subnetwork must be at
  least this dense.
\end{proof}

\begin{theorem}
  The density decomposition is unique. \label{thm:unique_rings}
\end{theorem}

\begin{proof}
  The maximum indegree of two egalitarian orientations for a given
  network is the same~\cite{bimowz2012,amoz2006,venkateswaran2004}.
  Suppose, for a contradiction, that there are two egalitarian
  orientations (red and blue) for $G$, resulting in density decompositions $R_0 ,R_1,
  \dots R_k$ and $B_0, B_1, \dots B_k$, respectively.  Let $i$ be the
  largest index such that $R_i \neq B_i$.

  \begin{figure}
    \centering
\includegraphics[scale = .13]{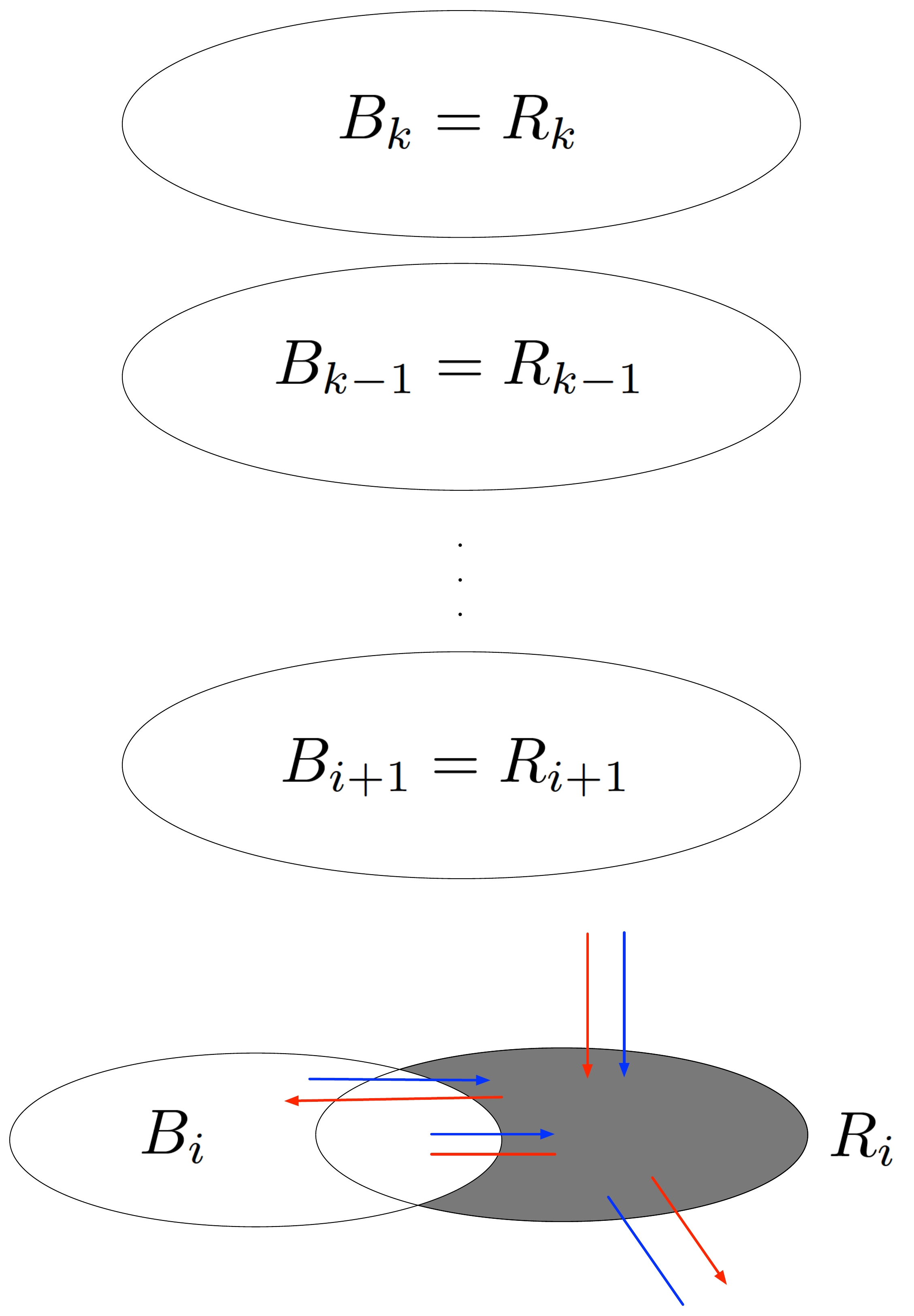}
\caption{Edges incident to $S$ (grey region) in the proof of Theorem
  \ref{thm:unique_rings} in red and blue orientations. $S$ is the
  region in gray.  Unoriented edges indicate that the orientation could
  be in either direction.  }
  \label{fig:illustrate_uniqueness}
  \end{figure}

  We compare the orientation of the edges with one endpoint in $S =
  R_i \setminus B_i$ between the two orientations (illustrated in
  Figure~\ref{fig:illustrate_uniqueness}).  Since the orientations are
  egalitarian:
  \begin{enumerate}
  \item All the edges between $B_i$ and $S$ are directed into $S$ in the blue
    orientation.
  \item All the edges between $S$ and $\{\cup_{j=i+1} ^k R_j\}\setminus
    S$ are directed into $S$ with respect to both red and blue orientations.
  \item All edges between $S$ and $\{\cup_{j=0} ^{i-1} R_j\} \setminus
    S$ are directed out of $S$ with respect to the red orientation.
  \end{enumerate}
  Based on these orientations, we have:
  \begin{observation}\label{obs:2}
    The number of edges directed into $S$ in the blue orientation is
    at least the number of edges directed into $S$ in the red
    orientation.
  \end{observation}

  We will show that $R_i \subseteq B_i$; symmetrically $B_i \subseteq
  R_i$, completing the theorem.

  With respect to the blue orientation, all nodes in $S$ have indegree
  strictly less than $i$. Further, by Observation~\ref{obs:2}, the
  total indegree shared amongst the nodes in $S$ with respect to the
  red orientation is at most that of the blue orientation. Since all
  nodes in $S$ have indegree $i$ or $i-1$ with respect to the red
  orientation, and, by Observation~\ref{obs:2}, the total indegree
  shared amongst the nodes in $S$ with respect to the red orientation
  is at most that of the blue orientation, all nodes in $S$ have
  indegree $i-1$ with respect to the red orientation.

  In order for every node in $S$ to have indegree $i-1$ in the red
  orientation, all nodes that are directed into $S$ in the blue
  orientation, must also be directed into $S$ in the red orientation;
  in particular this is true about the edges between $S$ and $R_i
  \setminus S$. Therefore, none of the nodes in $S$ (which have
  indegree $i-1$) reaches a node of $R_i\setminus S$ of indegree $i$
  with respect to the red orientation.  This contradicts the
  definition of $R_i$; therefore $S$ must be empty.
\end{proof}

The following theorem relies on the fact that the density decomposition is unique and proves Property~D\ref{prop:subring}.

\begin{theorem}
  \label{thm:densest_contained_in_top_ring}
  The densest subnetwork of a network $G$ is induced by a subset of the nodes in the densest ring of $G$.
\end{theorem}

\begin{proof}
  First note that the densest subnetwork is an induced subnetwork, for
  otherwise, the subnetwork would be avoiding including edges that
  would strictly increase the density.  Let $S$ be a set of nodes that
  induces a densest subnetwork of $G$.  Consider a density
  decomposition of $G$ and let $k$ be the maximum rank of a node in
  $G$.  Let $S_k = S \cap R_k$ and let $\bar S_k = S \setminus S_k$.

  Let $A$ be the set of edges in $G[S_k]$, let $C$ be the set of edges
  in $G[\bar S_k]$, and let $B$ be the edges of $G[S]$ that are
  neither in $G[S_k]$ or $G[\bar S_k]$.  We get
  \begin{equation}
    \label{eq:BC}
    |B|+|C| \le (k-1)|\bar S_k|
  \end{equation}
  because all the edges in $B$ and $C$ have endpoints in $\bar S_k$
  and all the nodes in $\bar S_k$ have indegree at most $k-1$ in
  the egalitarian orientation of $G$.
  \begin{equation}
    \label{eq:dH}
    \text{density}(G[S]) = \frac{|A|+|B|+|C|}{|S_k| + |\bar S_k|}
  \end{equation}
  \begin{eqnarray*}
    \text{density}(G[S_k]) &=& \frac{|A|}{|S_k|} \\
    &=& \frac{\text{density}(G[S])(|S_k| + |\bar S_k|)-(|B|+|C|)}{|S_k|} \qquad  \text{using Equation~(\ref{eq:dH})} \\
    &=& \text{density}(G[S]) + \frac{\text{density}(G[S])|\bar S_k|-(|B|+|C|)}{|S_k|} \\
    &\ge& \text{density}(G[S]) + \frac{\text{density}(G[S])|\bar S_k|-(k-1)|\bar S_k|}{|S_k|} \qquad \text{by Inequality~(\ref{eq:BC})}\\
    &>& \text{density}(G[S]) + \frac{(k-1)|\bar S_k|-(k-1)|\bar S_k|}{|S_k|} \qquad \text{by Lemma~\ref{lem:dense_between_k_k_1}}\\
    &=&  \text{density}(G[S])
  \end{eqnarray*}
  Therefore, removing the nodes of $G[S]$ that are not in $R_k$ produces a network of strictly greater density.
\end{proof}

\begin{theorem}
  \label{thm:indegree_clique}
For a clique on $n$ nodes, there is an orientation where each
node has indegree either $\lfloor n/2 \rfloor$ or $\lfloor n/2
\rfloor -1$.
\end{theorem}

\begin{proof}
  Give the nodes of the clique an ordering, $v_1 , v_2, \dots v_n$.
  Orient the edges between $v_1$ and $v_2 , \dots , v_{\lfloor n/2
    \rfloor +1}$ toward $v_1$ and edges between $v_1$ and $v_{\lfloor
    n/2 \rfloor +2},\dots v_n$ toward $v_{\lfloor n/2 \rfloor
    +2},\dots v_n$. Clearly $v_1$ has indegree $\lfloor n/2 \rfloor$.
  Similarly, for $v_2$: Orient the edges between $v_2$ and $v_3 ,
  \dots , v_{\lfloor n/2 \rfloor +2}$ toward $v_2$ and edges between
  $v_2$ and $v_{\lfloor n/2 \rfloor +3},\dots v_n$ toward $v_{\lfloor
    n/2 \rfloor +3},\dots v_n$. Clearly $v_2$ has indegree $\lfloor
  n/2 \rfloor$.  Continue in this fashion until $v_n$. It is immediate
  that $v_1,v_2,\dots v_{\lfloor n/2 \rfloor}$ have indegree $\lfloor
  n/2 \rfloor$.  Now for the remaining nodes: Consider $v_i$,
  ${\lfloor n/2 \rfloor} < i \leq n$. $v_i$ has $n-i$ incoming edges
  from nodes $v_{i+1}, \dots v_n$ and also ${i -\lfloor n/2 \rfloor
    -1}$ incoming edges from $v_1,\dots , v_{i -\lfloor n/2 \rfloor
    -1}$. Therefore $v_i$ has indegree $\lfloor n/2 \rfloor
  -1$. Therefore all nodes in the clique have indegree $\lfloor n/2
  \rfloor$ or $\lfloor n/2 \rfloor -1$.  Clearly such an orientation
  is egalitarian.
\end{proof}

\section{$k$-cores}\label{append:k-cores}

\begin{figure}
  \centering
\includegraphics[scale = .25]{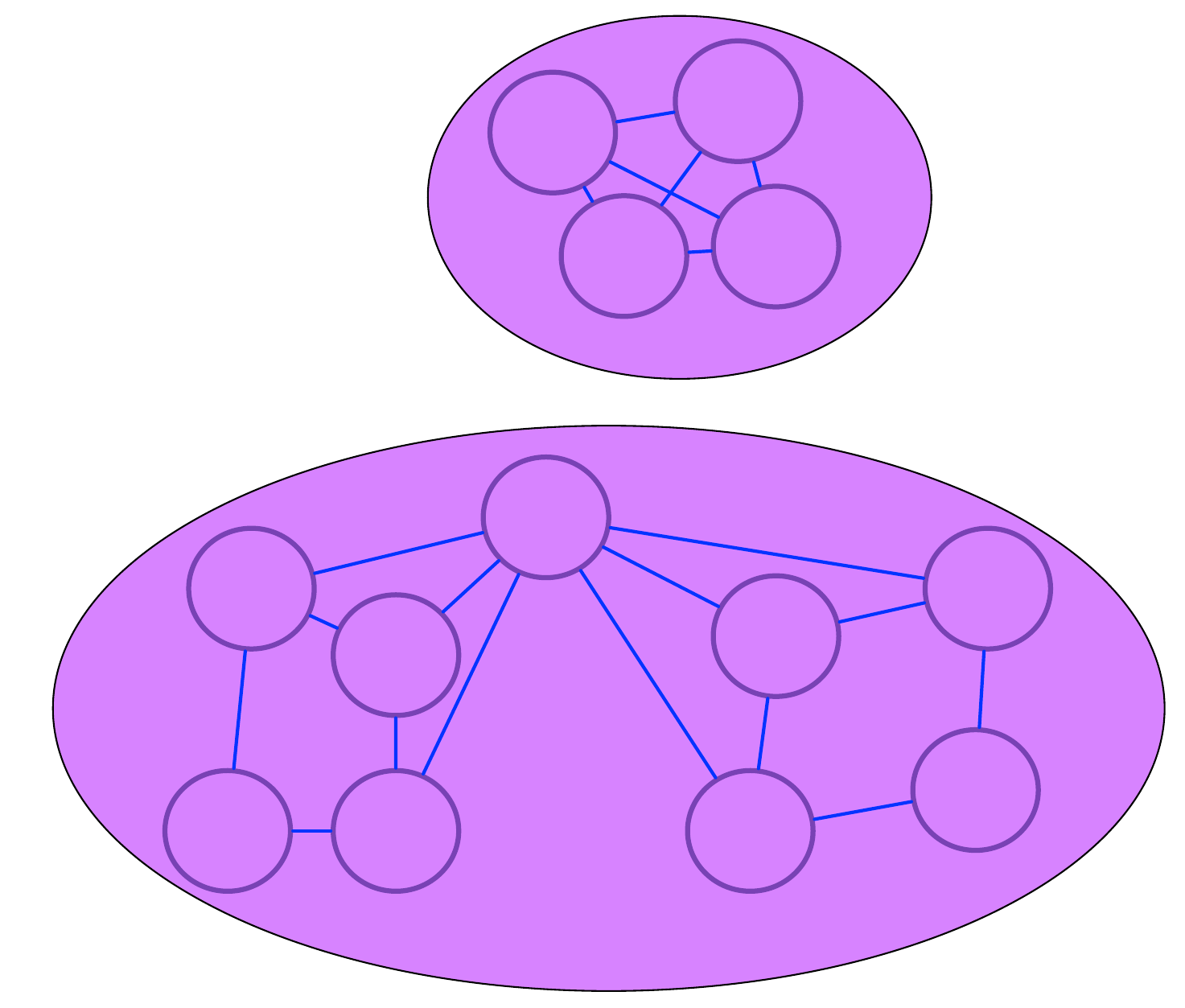}
  \caption{Core 3, the top core, has density 1.5. The densest subgraph
    is the subgraph induced by all vertices in core 2 and has
    density 1.56.}
  \label{fig:top_core_not_densest}
\end{figure}

See Figure~\ref{fig:top_core_not_densest} for an example of
a graph in which the densest subgraph is not contained in the top core. Further, while the core decomposition of a network can be
found in time linear in the number of edges~\cite{mb1983,bz2003,charikar2000} as opposed to the quadratic time required for
the density decomposition~\cite{bimowz2012}, core decompositions do
not lend themselves to a framework for building synthetic networks,
since it is not clear how to generate a $p$-core at random, whereas
density decompositions do.



\begin{figure}

\begin{minipage}{1\linewidth}
\centering
\subfloat[]{\label{main:a}\includegraphics[scale=.25]{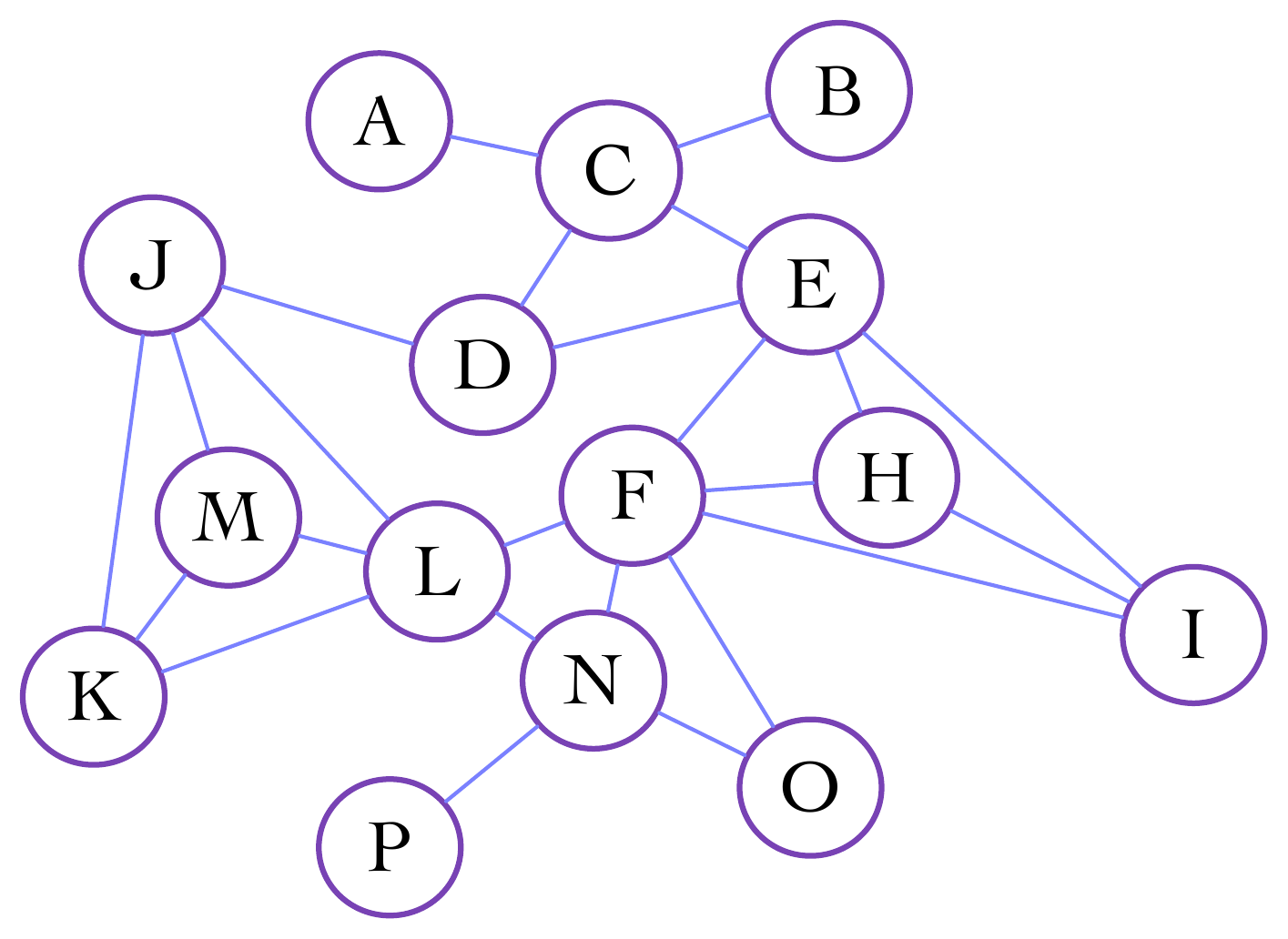}}
\end{minipage}
\par\medskip

\begin{minipage}{.5\linewidth}
\centering
\subfloat[]{\label{fig:ring_decomp_graph}\includegraphics[scale=.2]{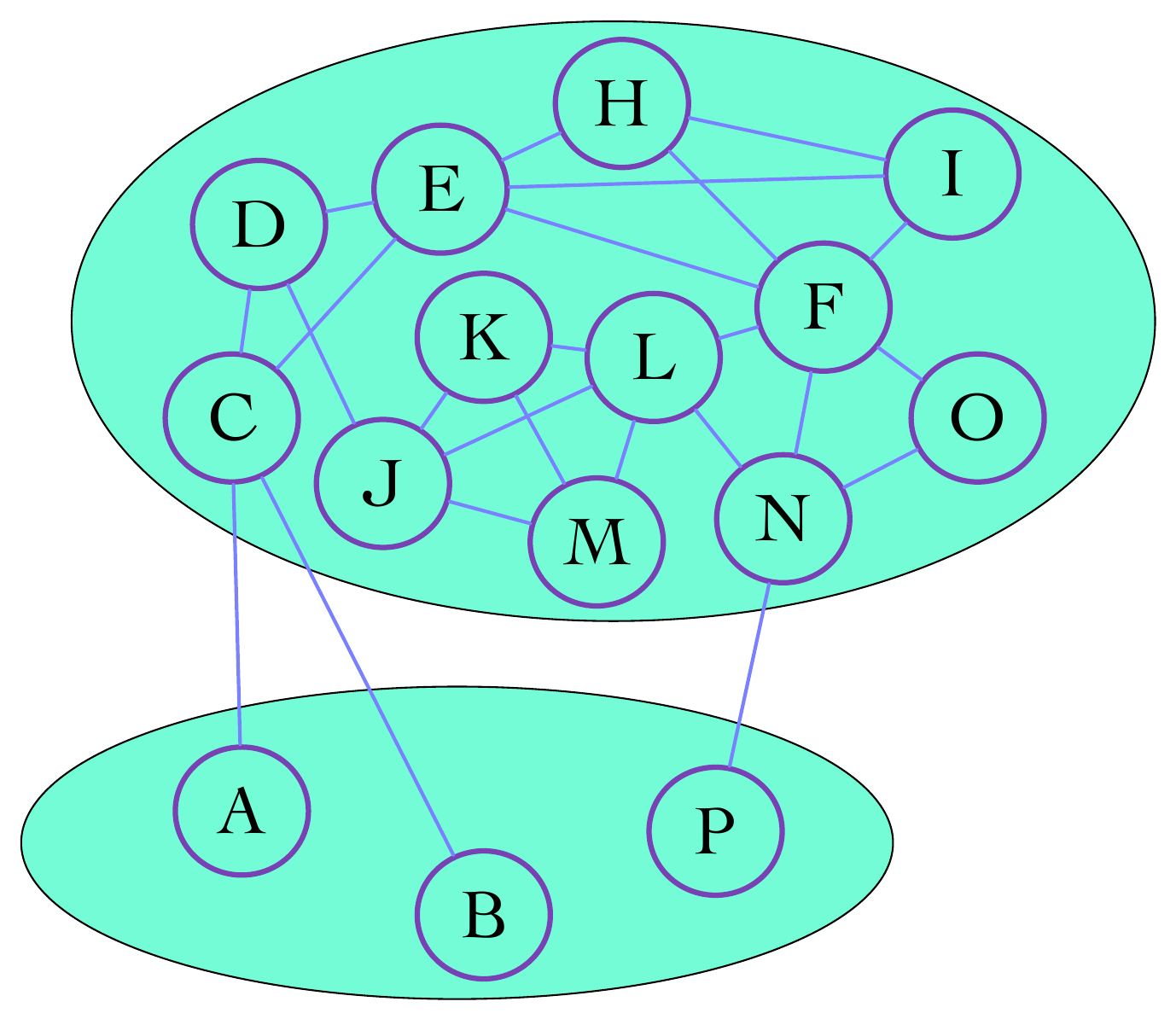}}
\end{minipage}
\begin{minipage}{.5\linewidth}
\centering
\subfloat[]{\label{fig:core_decomp_graph}\includegraphics[scale=.2]{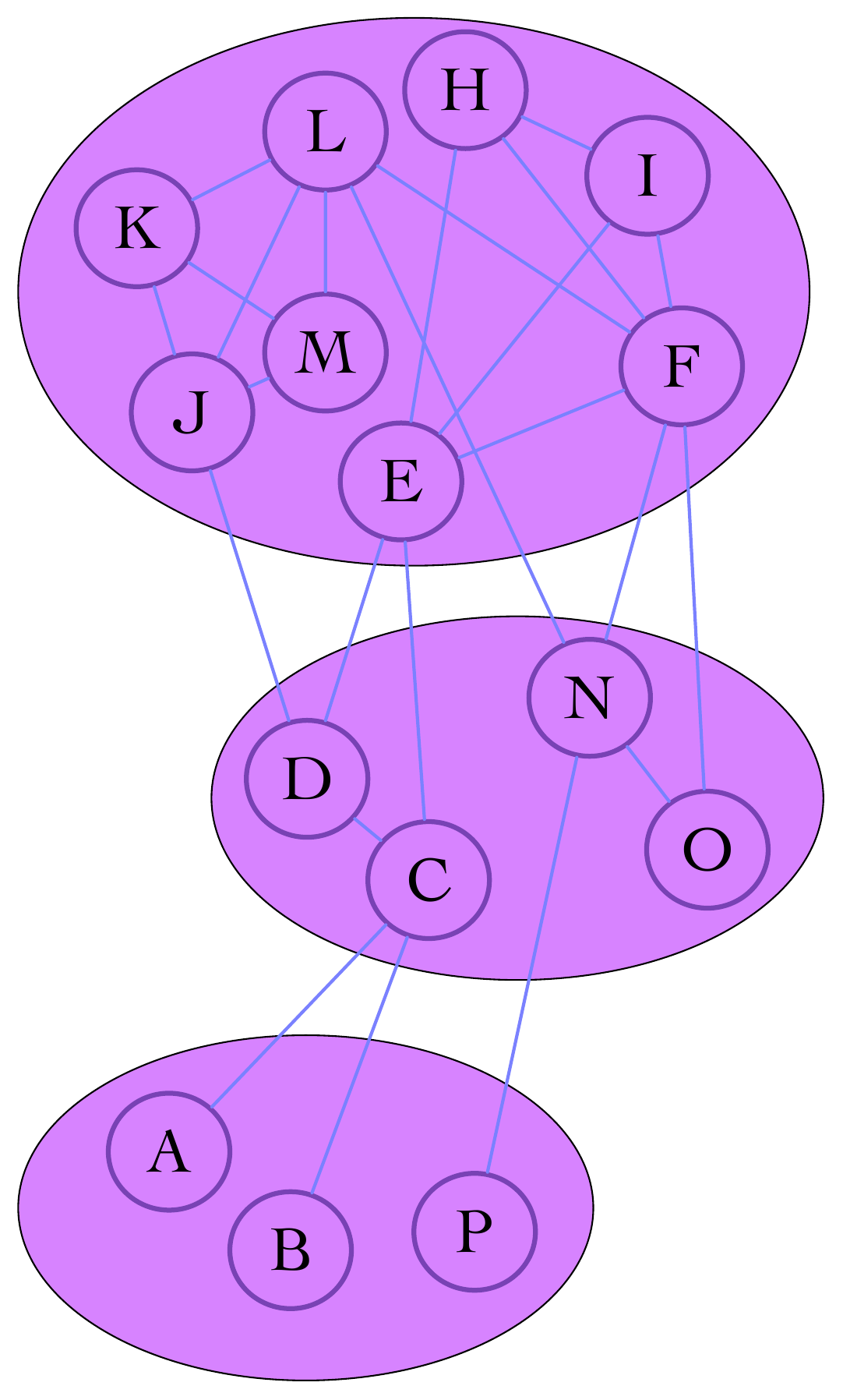}}
\end{minipage}

\caption[A graph illustrating that the subgraph induced by the
vertices of the top core is not necessarily as dense as the subgraph
induced by the vertices of the top ring.]{The top ring
  (Figure~\ref{fig:ring_decomp_graph}) contains
  vertices $I,K,L,M,C,D,E,H,F,J,N$ and $O$ but $C,D,N,$ and $O$ are
  not in the top core (Figure~\ref{fig:core_decomp_graph}). The density of the subgraph induced by the top
  ring is $21/12=1.75$ while the density of the subgraph induced by
  the top core is $13/8=1.625$.}
 \label{fig:cores_1}

\end{figure}

Recall that identifying the nodes in $\cup_{j > i} R_j$ and deleting
the nodes in $\cup_{j < i} R_j$ leaves a network $G$ whose density is
in the range $(i-1,i]$ (for $|R_i|$ sufficiently large). We find that
the bound on density for the corresponding cores is much
looser.

\begin{lemma}\label{lem:k-core-density}
  Given a core decomposition $H_0, H_1, \ldots, H_k$ of a network, the
  subnetwork formed by identifying the nodes in $\cup_{j > i}H_j$ and
  deleting the nodes in $\cup_{j< i} H_j$ has density in the range
  $[\frac{i}{2},i)$ for $|H_i|$ sufficiently large.
\end{lemma}

\begin{proof}
  Let $n$ be the number of nodes in the described subnetwork: $n =
  |H_i|+1$.  Let $d$ be the degree of the node resulting from the
  identification of $\cup_{j > i}H_j$.  Since every node in $H_i$ has
  degree at least $i$ in the subnetwork, the density of the subnetwork
  is at least ${{1\over 2}(i\cdot n+d)}\over n$, from which the lower
  bound of the lemma follows since $d \geq 0$.  This lower bound is also
  tight when $H_i$ induces an $i$-regular network.

  Further, the $i$-core is witnessed by iteratively deleting nodes of
  degree at most $i$ while such nodes exist.  The subnetwork will have
  the greatest density (the most edges) if each deletion removes a
  node of degree exactly $i$.  Then the subnetwork has density at
  most $i \cdot (n-1) \over n$.
\end{proof}

\begin{figure}[H]
  \centering
  \includegraphics[scale=0.5]{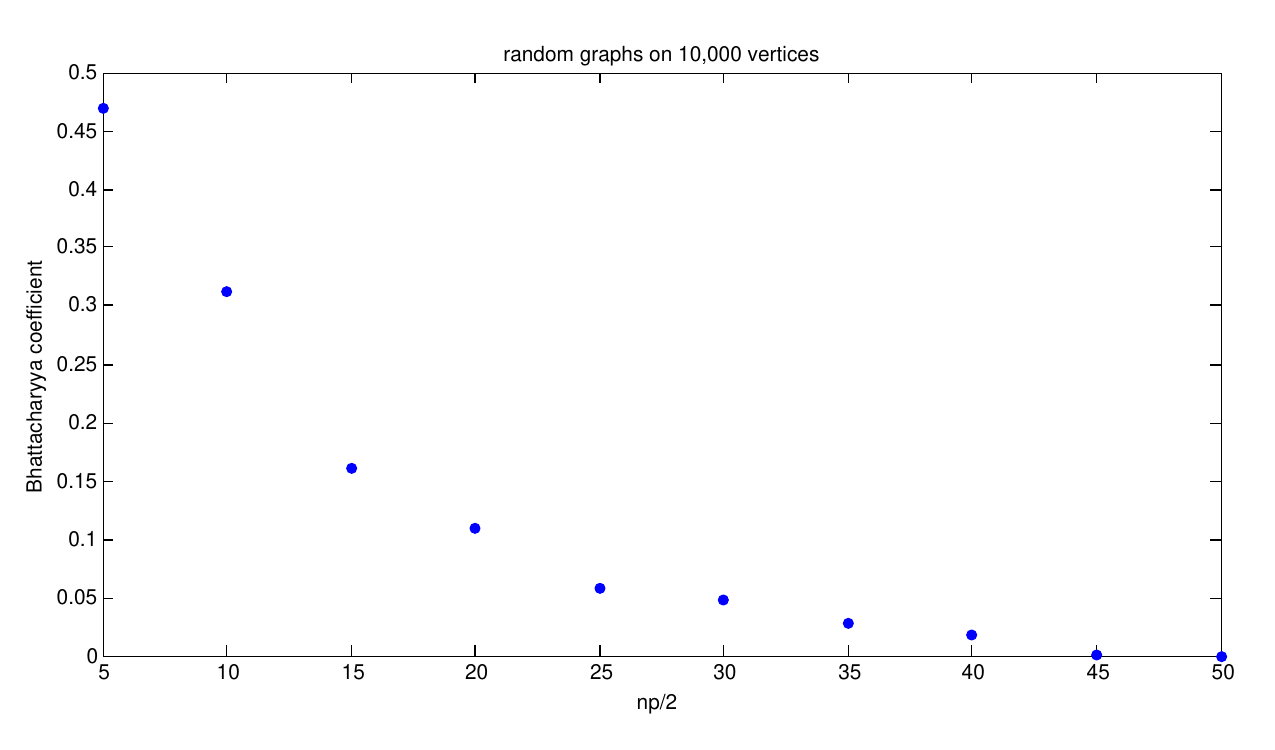}
  \caption{}
  \label{fig:bc_Gnp}
\end{figure} 

\section{Random networks with given density distributions}
\label{sec:model}

Based on the observation that real networks have similar density and
degree decompositions and that many synthetic networks have dissimilar
density and degree decompositions, we develop an abstract model, that,
given a particular density distribution, produces a network having
that density distribution (Section~\ref{sec:model}).  Applied
na\"ively, given a density distribution of a real network, this model
generates networks with realistic average path lengths (average number
of hops between pairs of nodes) and degree distributions; that is
similar to the given real network (Section~\ref{sec:rdd}).  In addition to
having short average path lengths, large-scale, real networks also tend
to have high {\em clustering coefficients}~\cite{newman2010}.  The
clustering coefficient of a node $v$ is the ratio of the number of
pairs of neighbors of $v$ that are connected to the number of pairs of
neighbors of $v$; the clustering coefficient of a network is the
average clustering coefficient of its nodes.  Our model, na\"ively
applied, unfortunately, but not surprisingly, results in networks with
very low clustering coefficients.  However, we show that
applying the abstract model in a more sophisticated manner, using
ideas from the small world model of Watts and Strogatz~\cite{ws1998},
results in much higher clustering coefficients (Section~\ref{sec:hsw})
suggesting that real networks may indeed be {\em hierarchies of small
  worlds}. 

Our hierarchies of small worlds specification is just one way to tune
our abstract model; our model is quite flexible allowing for the easy
incorporation of other network generation techniques, which we discuss
at the end of this paper.  A key observation that distinguishes our
model from other network models is our {\em qualitatively different}
treatment of nodes.  That is, our model begins by assigning nodes to
levels of the density decomposition.  This sets nodes qualitatively
apart from each other; for example, a node assigned to a dense level
of the decomposition is treated very differently from a node assigned
to a sparse level of the decomposition.

Unlike our model, other network models, such as the small worlds
model and the classic random graph model ($G_{n,p}$), treat each node in
the same way~\cite{ws1998}.  In the preferential attachment
model, in which nodes are attached one at a time to some fixed number
of existing nodes~\cite{BA99}, one may argue that nodes are treated
differently since they {\em arrive} to the network at different {\em
  times}.  However, when each node arrives, it is treated the same way
as nodes before it.  Similarly more recent network models, such as the
affiliation network~\cite{ls2009}, community-guided attachment and
forest-fire models~\cite{lkf2005}, nodes are not distinguished from
one another in a fixed way.  We posit that in order to generate
realistic networks, in particular networks exhibiting the rich density
hierarchy we have observed in all the networks we have tested, one
must assign nodes to classes and treat those classes
differently. There are existing models that treat nodes {\em
  qualitatively} differently as we do. In the fitness, or
Bianconi-Barab\'{a}si model, each node has an intrinsic value, or
``fitness''~\cite{bb2001}. As a network produced by this model
evolves, nodes with higher fitness are more likely to gain additional
edges. In the planted partition model the nodes of the graph are
divided into $\ell$ groups of size $\ell/|V|$, two nodes in the same
group are connected with probability $p$ and two nodes from different
groups are connected with probability $r<p$~\cite{js1998}. Thus nodes
in different groups are treated qualitatively differently.

\paragraph{Data sets} We note that our conclusions on the similarity
of density and degree distributions of a given network are stronger
for {\em self-determining} networks or those networks that represent
relationships, each of which is determined by at least one of the
parties in this relationship. Perhaps the clearest example of a
self-determining network is a social network in which nodes represent
people and an edge represents a friendship between two people.  On the
other hand a network representing the transformers and power lines
that connect them in a power grid is clearly not self-determining as
the transformers themselves do not determine which other transformers
they are connected to, but a power authority does.  For comparison, we
include two non-self-determining networks.

Table~\ref{tab:sdn} provides a list of the
real networks that we study.

\begin{table}[h]
  \centering
  \begin{tabular}[c]{l|l|l|l|l|c}
    \multicolumn{4}{c}{Self-determining networks}\\\hline
    Name & Nodes & $\#$ Nodes & Edges & $\#$ Edges & Source \\ \hline 
    AS & autonomous systems &44,729 & routing agreements & 170,735 & \cite{as_website}\\
    DBLP & computer scientists & 317,080 & at least one co-authored
                                           paper & 1,049,866 & \cite{yl2012} \\
    Enron & email addresses &36,692 & at least one email exchanged &
                                                                     183,831
                                                   &  \cite{ky2006} \\
    Epinions & {\tt epinions.com} members &75,879 & self-indicated
                                                    trust & 405,740 & \cite{tik2009}\\
    Facebook & Facebook user & 4,039 & Facebook friends & 88234&
                                                                 \cite{ml2014}  \\
    PHYS & condensed matter physicists & 40,421 & at least one co-authored paper& 175,692&  \cite{newman2004} \\
    Slashdot & {\tt slashdot.org} members & 82,168& indication of friend or foe & 504,230& \cite{lldm2008}\\
    Wikivote & {\tt wikipedia.org} users &7,115 & votes for administrator role & 103,689& \cite{lhk2012}\\
    \multicolumn{4}{c}{  }\\
    \multicolumn{4}{c}{Non-self-determining networks}\\\hline
    Name & Nodes& $\#$ Nodes & Edges & $\#$ Edges & Source \\ \hline 
    Amazon & products & 334,863 & pairs of frequently co-purchased items& 925,872& \cite{yl2012}\\
    Gnutella & network hosts & 22,687& connections for file sharing & 54,705& \cite{rfi2002}
  \end{tabular}
  \caption{Network data sets.  For naturally directed networks (Enron, Epinions and Wikivote), we ignore the directions and study the underlying undirected network.  We likewise ignore edge annotations (e.g.\ friend or foe in the Slashdot network).  We use three snapshots of the AS network (from 1999, 2005 and 2011) and three snapshots of the PHYS network (for papers posted to {\tt arxiv.org} prior to 1999, 2003 and 2005).  Note that the structure of the Gnutella network is given by external system design specifications.}
  \label{tab:sdn}
\end{table}

\subsection{Models}
Given a density distribution $\rhob$, we can generate a network with $n$ nodes having this density distribution using the following {\em abstract model}:
\begin{quote}
\begin{algorithmic}[1]
\INPUT density distribution \rhob\ and target size $n$
\OUTPUT a network $G$ with $n$ nodes and density distribution \rhob
\State Initialize $G$ to be a network with empty node set $V$
\For {$i =|\rhob|, \ldots, 0$} 
\State $R_i \gets $ set of $\lfloor \rho_i n \rfloor$  nodes
\State add $R_i$ to $V$
\For {\text{each node $v \in R_i$}}
\State connect $i$ nodes of $V$ to $v$ \label{alg:choice}
\EndFor
\EndFor
\end{algorithmic}
\end{quote}
Using this generic model, we propose two specific models, the {\em
  random density distribution model} (RDD) and the {\em hierarchical
  small worlds model} (HSW), by specifying how the neighbors are
selected in Step~\ref{alg:choice}.  First we show that this abstract
model does indeed generate a network with the given density
distribution:

\begin{lemma}
  \label{lem:dsm_egalitarian}
  The network resulting from the abstract model has density distribution $\rhob$.
\end{lemma}

\begin{proof}
  We argue that the orientation given by, in Step~\ref{alg:choice},
  directing the added edges into $v$ is egalitarian.  For a
  contradiction, suppose there is a reversible path.  There must be an
  edge on this path from a node $x$ to a node $y$ such that the
  in-degree of $y$ is strictly greater than the degree of $x$.  By
  construction, then, $x$ was added after $y$ and so an edge between
  $x$ and $y$ must oriented into $x$, contradicting the direction
  required by the reversible path.

  Finally, since the nodes in set $R_i$ have indegree $i$ according to
  this orientation, the orientation is a witness to a density
  decomposition of the given distribution.
\end{proof}

Notice that in this construction, nodes in $R_i$ will have indegree
$i$ while a network with the same density decomposition may have nodes
in $R_i$ with indegree $i-1$. We could additionally specify the number
of nodes in $R_i$ that have indegree $i$ and indegree $i-1$; this
would additionally require ensuring that there is an egalitarian
orientation in which all the nodes destined to have indegree $i-1$ in
$R_i$ {\em reach} nodes of indegree $i$ in $R_i$.  We believe this is
needlessly over-complicated and, indeed, over-specification that will
have little affect on the generation {\em large} realistic networks.

Further notice that this abstract model may generate a network
that is not simple.  Without further constraint, in
Step~\ref{alg:choice}, $v$ may connect to itself (introducing a
self-loop) or to a node that $v$ is already connected to (introducing
parallel edges).  We adopt a simple technique used for generating
$d$-regular networks~\cite{mw1990}: we constrain the choice in
Step~\ref{alg:choice} to nodes of $V$ that are not $v$ itself nor
neighbors of $v$.  McKay and Wormald prove this constraint still allows for
uniformity of sampling of $d$-regular networks when $d$ is sufficiently small ($d =
O(n^{1/3})$)~\cite{kv2003}; likewise, since $i$ is small compared to
$|R_i|$ for large networks, adopting this technique should not affect
our sampling.  In our two specific models, described below, we ensure
the final network will be simple using this technique.

\subsection{Random density distribution model}\label{sec:rdd}

\begin{figure}[h]
  \centering
  \includegraphics[scale=0.7]{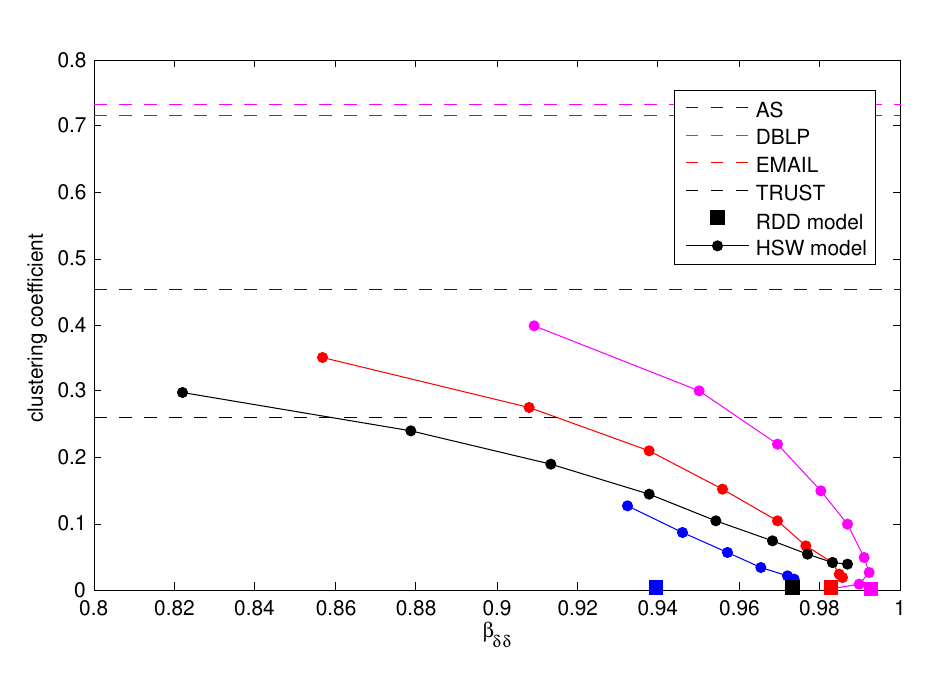}
  \caption{Clustering coefficient versus similarity of degree
    distribution (between models and original networks, $\beta
    _{\delta \delta}$) for RDD and HSW models.  Measurements for
    the SW model networks are not shown as $\beta_{\delta\delta} <
    0.4$ for all networks generated.   Dotted lines represent the clustering
    coefficients for the original networks.  }
  \label{fig:cc_vs_bhatta_delta_delta}
\end{figure}

\begin{figure}[h]
  \centering
  \includegraphics[scale=0.7]{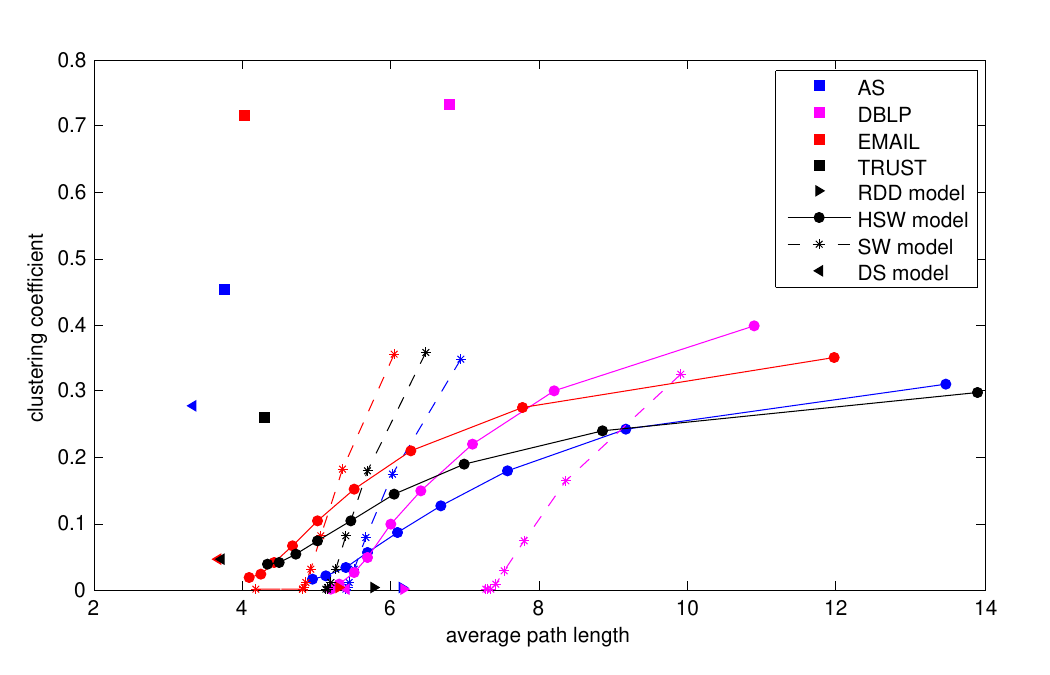}
  \caption{Clustering coefficient versus average path length for RDD,
    HSW, SW and DS models. Colors indicate the network being modelled.  Squares denote the data for the original networks.}
  \label{fig:cc_vs_apl}
\end{figure}

For the RDD model, we choose $i$ nodes from $V$ uniformly at random in
Step~\ref{alg:choice}.  We use this to model four networks in our data
set (AS, DBLP, EMAIL, and TRUST).  For each given network, we generate
another random network having the given network's number of nodes and
density distribution.  Remarkably, although we are only specifying the distribution of the nodes over a density decomposition, the resulting degree distributions of the RDD networks are very similar to the original networks they are modeling.  We use the Bhattacharyya coefficient to quantify the similarity between the normalized degree distribution of an RDD network and the normalized degree distribution of the original network; we denote this by $\beta_{\delta\delta}$ (to distinguish from our use of the Bhattacharyya coefficient to compare degree distributions to density distributions.  For all four models, 
$\beta_{\delta \delta}>0.93$ (Figure~\ref{fig:cc_vs_bhatta_delta_delta}).  Further, the average path lengths of the RDD networks are realistic, within ~2 of the average path lengths of the original networks (Figure~\ref{fig:cc_vs_apl}).

\begin{figure}[h]
  \centering
  \includegraphics[scale=0.7]{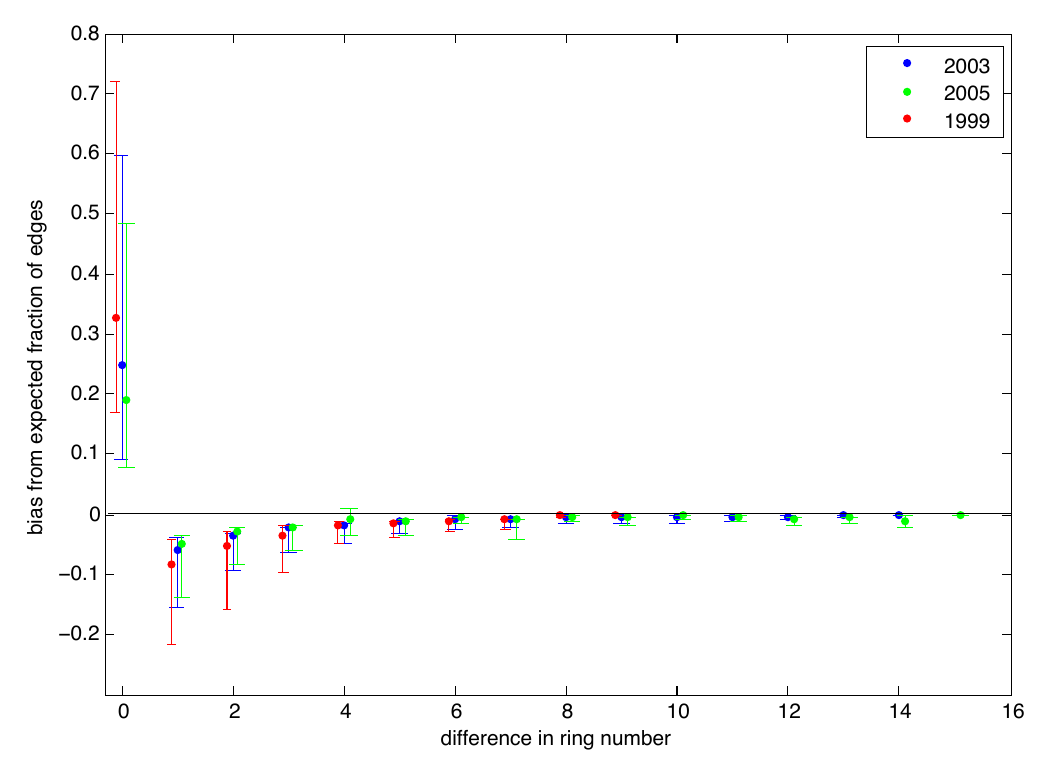}
  \caption{Range of difference between the actual fraction of edges connecting $R_i$ to $R_j$ and expected fraction (Equation~(\ref{eq:frac})) over all $j \ge i$ as a function of $j-i$ for three PHYS networks.  Error bars show max and min values of these
    differences and dots indicate the average.}
  \label{fig:edge-bias}
\end{figure}

However, the clustering coefficients of the RDD networks are
unrealistically low (Figures~\ref{fig:cc_vs_bhatta_delta_delta}
and~\ref{fig:cc_vs_apl}).  Upon further inspection, we find that, for
example, the PHYS networks have many more edges between nodes of a
common ring of its density decomposition than between rings as
compared to the corresponding RDD model.  For the RDD model, we can
compute the expected fraction of edges that will have one endpoint in
$R_i$ and one endpoint in $R_j$.  Since there are $|R_j||R_i|$ such
edges to choose from (for $j > i$) and at most
$|R_i|(\frac{1}{2}(|R_i|-1)+\sum_{j > i} |R_j|))$ edges between $R_i$
and $R_j$ (for $j > i$), we would expect this fraction to be:
\begin{equation}\label{eq:frac}
  \frac{|R_j|}{\frac{1}{2}(|R_i|-1)+\sum_{j > i} |R_j|} \text{ for $j > i$ and }  \frac{\frac{1}{2}(|R_i|-1)}{\frac{1}{2}(|R_i|-1)+\sum_{j > i}
  |R_j|}\text{ for $i= j$}
\end{equation}
In Figure~\ref{fig:edge-bias} we plot the
difference between the actual fraction of edges connecting $R_i$ to
$R_j$ in the PHYS networks with this expected fraction for all values
of $j-i$. We see that when $j-i = 0$, or for edges with both endpoints
in the same ring, there is a substantially larger number of edges in
the original networks than is being captured by our model.  This
provides one explanation for the low clustering coefficients produced
by the RDD model.

\subsection{Hierarchical small worlds model}\label{sec:hsw}

We provide a more sophisticated model which addresses the
unrealistically low clustering coefficients of the RDD model by
generating a small world (SW) network among the nodes of each ring of
the density decomposition.  Recall that a SW network on $n$ nodes,
average degree $d$ and randomization $p$ network is created as
follows: order the nodes cyclically and connect each node to the $d$
nodes prior to it; with probability $p$ reconnect one endpoint of each
edge to another node chosen uniformly at random.  The SW model
provides a trade-off between clustering coefficient and average path
length: as $p$ increases, the clustering coefficient and the average
path length decreases~\cite{ws1998}.

In the hierarchical small worlds (HSW) model, for nodes in $R_i$, we
create a SW network on $|R_i|$ nodes and average degree $i$ in the
same way, except for how we reconnect each edge with probability $p$.
For an edge $uv$ where $u$ is a node within $d$ nodes prior to $v$ in
the cyclic order, we select a node $x$ uniformly at random from
$\cup_{j>i} R_j$ and replace $uv$ with $xv$.  For the densest ring, we
select a node uniformly at random from the densest ring.  

This process is exactly equivalent to the following: order $R_i$
cyclically; for each $v \in R_i$, with probability $p$, connect each
of the $i$ nodes before $v$ in this order to $v$; if $c \le i$
neighbors for $v$ are selected in this way, select $i-c$ nodes
uniformly at random from $\cup_{j>i}R_j$ (or $R_i$ if this is the
densest ring) and connect these to $v$.  Clearly, this is a
specification of neighbor selection for Step~\ref{alg:choice} of the
abstract model.

For the AS, DBLP, EMAIL, and TRUST networks in our data set, we
generate a random network according to the HSW model that is of the
same size and density distribution of the original network.  We do so
for $p = 0.1, 0.2, \ldots, 0.9$.  As with the SW model, the HSW model
provides a similar trade-off between clustering coefficient and
average path length (Figure~\ref{fig:cc_vs_apl}), although the
relationship is less strong.  In addition, we observe a similar trade-off between $p$ and degree distribution: as $p$ increases, the degree distribution approaches that of the original network (Figure~\ref{fig:cc_vs_bhatta_delta_delta}).  This is in sharp contrast to the SW model which have degree distributions far from the original (normal vs.\ close to power law).

\subsection{Comparing to the degree sequence model} \label{sec:ds}

We also compare our models (RDD and HSW) to a {\em degree sequence}
(DS) model.  For a given degree distribution or sequence (assignment
of degree to each node), a DS model will generate a graph, randomly,
having that degree sequence.  We use the model of Viger and Latapy
which generates a connected, simple graph by iteratively selecting neighbors for nodes (from highest remaining degree to be satisfied to lowest) and randomly shuffling to prevent the process from getting stuck
(if no new neighbor exists that has not yet fulfilled its prescribed degree)~\cite{vl2005}.  As with RDD and HSW
we generate a network using this DS model corresponding to the degree
sequence of the AS, EMAIL, and TRUST networks.  The clustering
coefficients of the resulting networks are much lower than in the real
networks (Figure~\ref{fig:cc_vs_apl}); in the case of the AS network,
this mismatch is less extreme, most likely because this network has an
extremely long tail with a node with degree 4,171; many nodes would
connect to these high degree nodes, providing an opportunity for
clustering. The average path lengths are close to the original
networks.  Notably, the density distributions of the networks
generated by the DS model are very similar to their degree
distribution, all having $\beta_{\rho\delta} > 0.9$.

These observations for the DS model add evidence to our proposal that
in order to generate realistic networks, one must distinguish between
types of nodes; doing so results in networks that resemble real
networks.  However, we must note that the DS model suffers from two
drawbacks.  First, the algorithms for generating such networks are
much less efficient than our models (RDD and HSW, which run in linear
time); in order to guarantee simplicity and connectivity, the
reshuffling required incurs a large computational overhead,
particularly when the degree sequence includes very high degree nodes
(such as in the AS network).  Second, the amount of information
required to specify network generation via the DS model is much greater
than our abstract model.  In the former, the degree of every node must
be specified, or at least the number of nodes having each degree.  For
example, the SLASH network has 457 unique degrees (and a maximum
degree of 2553) while only having 61 non-empty rings in the density
decomposition.

\end{document}